%% file: n2hdm.tex
\pdfoutput=1
\documentclass[12pt]{article}
\usepackage{graphicx}
\usepackage{subcaption}
\usepackage{amssymb}
\usepackage{amsmath}
\usepackage{amsfonts}

\usepackage{xcolor}
\usepackage{url}
\usepackage{float}
\usepackage{float}
\usepackage{jheppub}

\usepackage{pifont}% http://ctan.org/pkg/pifont
\newcommand{\cmark}{\ding{51}}%
\newcommand{\xmark}{\ding{55}}%

\usepackage{soul}
\usepackage[english]{babel}
\usepackage[T1]{fontenc}
\usepackage{lmodern}
\usepackage[utf8]{inputenc}  % f??r Unix-Systeme
\usepackage{bbm}

\usepackage{hyperref}
\hypersetup{
  colorlinks   = true, %Colours links instead of ugly boxes
  urlcolor     = blue, %Colour for external hyperlinks
  linkcolor    = black, %Colour of internal links
  citecolor   = black %Colour of citations
}

\usepackage[numbers,sort&compress]{natbib}
\graphicspath{{plots/}}

\input{paperdef.tex}

\title{
A 96 GeV Higgs Boson in the N2HDM
}
\author[1,2]{T.~Biek\"otter,}
\author[1]{M.~Chakraborti,}
\author[1,3,4]{S.Heinemeyer}
\affiliation[1]{
Instituto de F\'isica Te\'orica UAM-CSIC, Cantoblanco, 28049, Madrid, Spain}

\affiliation[2]{Departamento de F{\'i}sica Te{\'o}rica, Universidad Aut{\'o}noma
de Madrid (UAM), 
Campus de Cantoblanco, 28049 Madrid, Spain}

\affiliation[3]{
Campus of International Excellence UAM+CSIC,
Cantoblanco, 28049, Madrid, Spain}

\affiliation[4]{Instituto de F\'isica de Cantabria (CSIC-UC),
39005, Santander, Spain}

\emailAdd{thomas.biekotter@csic.es}
\emailAdd{mani.chakraborti@gmail.com}
\emailAdd{Sven.Heinemeyer@cern.ch}
\abstract{
We discuss a $\sim 3\,\sigma$
signal (local) in the light Higgs-boson search
in the diphoton decay mode at $\sim 96 \gev$ as reported by
CMS, together with a $\sim 2\,\sigma$ excess (local)
in the $b \bar b$ final state
at LEP in the same mass range.
We interpret this possible signal as a Higgs boson in the
2~Higgs Doublet Model with an additional real Higgs singlet (N2HDM).
We find that the lightest Higgs boson of the N2HDM can perfectly fit
both excesses simultaneously, while the second lightest state is
in full agreement with the Higgs-boson measurements at
$125 \gev$, and the full Higgs-boson
sector is in agreement with all Higgs exclusion
bounds from LEP, the Tevatron and the LHC as well as other theoretical and
experimental constraints. We show that only the N2HDM type~II and~IV can
fit both the LEP excess and the CMS excess
with a large ggF production component at $\sim 96 \gev$.
We derive bounds on the N2HDM Higgs sector from a fit to both excesses
and describe how this signal can be further analyzed at
the LHC and at future $e^+e^-$~colliders, such as the ILC.
}

\preprint{IFT-UAM/CSIC--19--034\\ \mbox{}\hfill arXiv:1903.11661}

\begin{document}
\maketitle

\section {Introduction}
\label{sec:intro}

In the year 2012 the ATLAS and CMS collaborations have discovered a new
particle that -- within theoretical and experimental uncertainties -- is
consistent with the existence of a Standard-Model~(SM) Higgs boson at a mass
of~$\sim 125 \gev$~\cite{Aad:2012tfa,Chatrchyan:2012xdj,Khachatryan:2016vau}.
No conclusive signs of physics beyond the~SM have been found so far at the LHC.
However, the measurements of Higgs-boson couplings, which are known
experimentally to a precision of roughly $\sim 20\%$, leave room for
Beyond Standard-Model (BSM) interpretations. Many BSM models possess
extended Higgs-boson sectors. Consequently, one of the main tasks of the
LHC Run~II and beyond is to determine whether 
the observed scalar boson forms part of the Higgs sector of an extended
model.
Extended Higgs-boson sectors naturally contain additional Higgs bosons with
masses larger than $125 \gev$. However, many extensions also offer the
possibilty of additional Higgs bosons {\em lighter} than
$125 \gev$. Consequently, the search for lighter Higgs bosons forms an
important part in the BSM Higgs-boson analyses.

Searches for Higgs bosons below $125 \gev$ have been performed at LEP,
the Tevatron and the LHC. 
LEP reported a $2.3\,\sigma$ local excess
observed in the~$e^+e^-\to Z(H\to b\bar{b})$
searches\,\cite{Barate:2003sz}, which would be consistent with a
scalar of mass ~$\sim 98 \gev$, but due to the $b \bar b$ final state the 
mass resolution is rather coarse). The excess corresponds to 
\begin{equation}
\mu_{\rm LEP}=\frac{\sigma\left( e^+e^- \to Z \phi \to Zb\bar{b} \right)}
			   {\sigma^{\mathrm{SM}}\left( e^+e^- \to Z H
			   		\to Zb\bar{b} \right)}
			  = 0.117 \pm 0.057 \; ,
\label{muLEP}
\end{equation}
where the signal strength $\mu_{\rm LEP}$ is the
measured cross section normalized to the SM expectation,
with the SM Higgs-boson mass at $\sim 98\gev$.
The value for $\mu_{\rm LEP}$ was extracted in \citere{Cao:2016uwt}
using methods described in \citere{Azatov:2012bz}.

Interestingly, recent CMS~Run\,II
results\,\cite{Sirunyan:2018aui} for Higgs-boson searches in the diphoton
final state show a local excess of~$\sim 3\,\sigma$ around
$\sim 96 \gev$, with a similar excess
of~$2\,\sigma$ in the Run\,I data at a comparable
mass~\cite{CMS:2015ocq}.
In this case the excess corresponds to (combining 7, 8 and $13 \tev$
data, and assuming that the $gg$ production dominates)
\begin{equation}
\mu_{\rm CMS}=\frac{\sigma\left( gg \to \phi \to \gamma\gamma \right)}
         {\sigma^{\rm SM}\left( gg \to H \to \gamma\gamma \right)}
     = 0.6 \pm 0.2 \; .
\label{muCMS}
\end{equation}
First Run\,II~results from~ATLAS
with~$80$\,fb$^{-1}$ in the~$\ga\ga$~searches below~$125$\,GeV were
recently published~\cite{ATLAS:2018xad}. No significant excess above
the~SM~expectation was observed in the mass range between $65$~and
$110 \gev$. 
However, the limit on cross section times branching ratio obtained in the
diphoton final state by ATLAS is not only well above $\mu_{\rm CMS}$,
but even weaker than the corresponding upper limit obtained by CMS at 
$\sim 96 \gev$. This was illustrated in Fig.~1
in \citere{Heinemeyer:2018wzl}. 

Since the CMS and the LEP excesses in the light Higgs-boson searches
are found effectively at the same mass, this 
gives rise to the question whether they might be of a common origin -- and 
if there exists a model which could explain the two excesses
simultaneously, while being in agreement with all other Higgs-boson
related limits and measurements. A review about
these possibilities was given
in \citeres{Heinemeyer:2018jcd,Heinemeyer:2018wzl}. The list comprises of
type~I 2HDMs~\cite{Fox:2017uwr,Haisch:2017gql}, a radion
model~\cite{Richard:2017kot},
a minimal dilaton model~\cite{LiuLiJia:2019kye},
as well as supersymmetric
models~\cite{Biekotter:2017xmf,Domingo:2018uim,Hollik:2018yek}.

\medskip
Motivated by the Hierarchy Problem, Supersymmetric
extensions of the~SM play a prominent role in the exploration of new
physics. 
Supersymmetry~(SUSY) doubles the particle degrees of freedom by predicting
two scalar partners for all SM fermions, as well as fermionic partners
to all SM bosons. The simplest SUSY extension of the SM is the Minimal Supersymmetric
Standard Model (MSSM)~\cite{Nilles:1983ge,Haber:1984rc}.
In contrast to the single Higgs doublet of the SM, the MSSM by construction,
requires the presence of two Higgs doublets, $\Phi_1$ and $\Phi_2$. 
In the $\cp$ conserving case the MSSM Higgs sector 
consists of two $\cp$-even, one
$\cp$-odd and two charged Higgs bosons. The light (or the heavy)
$\cp$-even MSSM Higgs boson can be interpreted as the signal discovered
at $\sim 125 \gev$~\cite{Heinemeyer:2011aa} (see
\citeres{Bechtle:2016kui,Bahl:2018zmf} for recent updates). However,
in \citere{Bechtle:2016kui} it was demonstrated that the MSSM cannot
explain the CMS excess in the diphoton final state. 
   
Going beyond the MSSM, a well-motivated extension is given by
the Next-to-MSSM (NMSSM) (see
~\cite{Ellwanger:2009dp,Maniatis:2009re} for reviews). 
The NMSSM provides a solution for the so-called ``$\mu$ problem'' by
naturally associating an adequate scale to the $\mu$ parameter appearing
in the MSSM superpotential~\cite{Ellis:1988er,Miller:2003ay}.
In the NMSSM a new singlet superfield is introduced, which only couples to the
Higgs- and sfermion-sectors, giving rise to an effective $\mu$-term,
proportional to the vacuum expectation value (vev) of the scalar singlet.
In the $\cp$ conserving case the NMSSM Higgs sector consists of
three $\cp$-even Higgs bosons,
$h_i$ ($i = 1,2,3$), two $\cp$-odd Higgs bosons, $a_j$ ($j = 1,2$),
and the charged Higgs boson pair $H^\pm$. 
In the NMSSM not only the
lightest but also the second lightest $\cp$-even Higgs boson
can be interpreted as the signal observed at about $125~\gev$, see,
e.g., \cite{King:2012is,Domingo:2015eea}.
In \citere{Domingo:2018uim} it was found that the NMSSM can indeed
simultaeneously satisfy the two excesses mentioned above.
In this case, the Higgs boson at $\sim 96 \gev$ has a large 
singlet component, but also a sufficiently large doublet 
component to give rise to the two excesses. 

A natural extension of the NMSSM is the \mnSSM, in which the 
singlet superfield is interpreted as a right-handed neutrino 
superfield~\cite{LopezFogliani:2005yw,Escudero:2008jg} 
(see \citeres{Munoz:2009an,Munoz:2016vaa,Ghosh:2017yeh} for reviews). 
The \mnSSM\ is the simplest extension of the MSSM that can provide 
massive neutrinos through a see-saw mechanism at the electroweak scale.
A Yukawa coupling for right-handed neutrinos of the order of the
electron Yukawa coupling is introduced that induces the explicit
breaking of $R$-parity. 
Also in the \mnSSM\ the signal at $\sim 125 \gev$ can
be interpreted as the lightest or the second lightest $\cp$-even scalar.
In \citere{Biekotter:2017xmf} the ``one generation case'' (only one
generation of massive neutrinos) was analyzed: within the scalar sector,
due to $R$-parity breaking, the left- and right-handed sneutrinos  
mix with the doublet Higgses and form six massive $\cp$-even 
and five massive $\cp$-odd states, assuming that there is no 
$\cp$-violation.
However, due to the smallness of $R$-parity breaking in the \mnSSM,
the mixing of the doublet Higgses with the left-handed
sneutrinos is very small.
Consequently, in the one-generation case the %scalar
Higgs-boson sector of the \mnSSM, i.e. the $\cp$-even/odd Higgs doublets
and the $\cp$-even/odd right handed sneutrino, resembles
the Higgs-boson sector in the NMSSM. 
In \citere{Biekotter:2017xmf} it was found that also the \mnSSM\
can fit the CMS and the LEP excesses simultaneously. In this case the
scalar at $\sim 96 \gev$ has a large right-handed sneutrino component.
The three generation case (i.e.\ with three generations of massive
neutrinos) is currently under investigation~\cite{munuSSM-3g}.

Motivated by the fact that two models with two Higgs doublets and
(effectively) one Higgs singlet can fit the CMS excess in \refeq{muCMS}
and the LEP excess in \refeq{muLEP}, we investigate in this work the 
Next to minimal two Higgs doublet model
(N2HDM)~\cite{Chen:2013jvg,Muhlleitner:2016mzt}.
Similar to the above two models, in N2HDM the two Higgs doublets are
supplemented with a real Higgs singlet, giving rise 
to one additional (potentially light) $\cp$-even Higgs boson. However, 
in comparison with the NMSSM and the \mnSSM\ the N2HDM does not have to
obey the SUSY relations in the Higgs-boson sector. Consequently, it
allows to study how the potential fits the two excesses
simultaneously in a more general way. Our paper is organized as follows. 
In \refse{sec:model} we describe the relevant features of the N2HDM. The
experimental and theoretical constraints taken into account are given
in \refse{sec:constraints}. Details about the experimental excesses as
well as how we implement them are summarized
in \refse{sec:excesses}. In \refse{sec:results} we show our results in
the different versions of the N2HDM and discuss the possibilities to
investigate these scenarios at current and future colliders. We conclude
with \refse{sec:conclusion}.

%%%%%%%%%%%%%%%%%%%%%%%%%%%%%%%%%%%%%%%%%%%%%%%%%%%%%%%%%%%%%%%%%%%%
%%%%%%%%%%%%%%%%%%%%%%%%%%%%%%%%%%%%%%%%%%%%%%%%%%%%%%%%%%%%%%%%%%%%

\section {The model: N2HDM}
\label{sec:model}

The N2HDM 
is the simplest 
extension of a $\cp$-conserving two Higgs doublet model (2HDM) 
in which the latter is augmented with a real 
scalar singlet Higgs field. The scalar potential of this model can be given 
as~\cite{Chen:2013jvg,Muhlleitner:2016mzt}
\begin{eqnarray}
V &=& m_{11}^2 |\Phi_1|^2 + m_{22}^2 |\Phi_2|^2 - m_{12}^2 (\Phi_1^\dagger
\Phi_2 + h.c.) + \frac{\lambda_1}{2} (\Phi_1^\dagger \Phi_1)^2 +
\frac{\lambda_2}{2} (\Phi_2^\dagger \Phi_2)^2 \nonumber \\
&& + \lambda_3
(\Phi_1^\dagger \Phi_1) (\Phi_2^\dagger \Phi_2) + \lambda_4
(\Phi_1^\dagger \Phi_2) (\Phi_2^\dagger \Phi_1) + \frac{\lambda_5}{2}
[(\Phi_1^\dagger \Phi_2)^2 + h.c.] \nonumber \\
&& + \frac{1}{2} m_S^2 \Phi_S^2 + \frac{\lambda_6}{8} \Phi_S^4 +
\frac{\lambda_7}{2} (\Phi_1^\dagger \Phi_1) \Phi_S^2 +
\frac{\lambda_8}{2} (\Phi_2^\dagger \Phi_2) \Phi_S^2 \;,
\label{eq:scalarpot}
\end{eqnarray}
\noindent
where $\Phi_1$ and $\Phi_2$ are the two $SU(2)_L$ doublets whereas
$\Phi_S$ is a real scalar singlet. 
To avoid the occurrence of tree-level flavor
changing neutral currents (FCNC), a $Z_2$ symmetry is imposed on the 
scalar potential of the model under which the 
scalar fields transform as
\begin{align}
  \Phi_1 \to \Phi_1\;, \quad \Phi_2 \to - \Phi_2\;, \quad
  \Phi_S \to \Phi_S~.
  \label{eq:2HDMZ2}
\end{align}
This $Z_2$, however, is softly broken by the $m_{12}^2$ term in
the Lagrangian. The extension of the $Z_2$ symmetry to the Yukawa
sector forbids
tree-level FCNCs. 
As in the 2HDM, one can have four variants of 
the N2HDM, depending on the $Z_2$ parities of the 
fermions. \refta{tabtypes} lists the couplings 
for each type of fermion 
allowed by the $Z_2$ parity in four different types of N2HDM.

%%%%%%%%%%%%%%%%%%%%%%%%%% TABLE %%%%%%%%%%%%%%%%%%%%%%%%%%%%%%%%
\begin{table}[t!]
\begin{center}
\begin{tabular}{lccc} 
\hline
  & $u$-type & $d$-type & leptons \\
\hline
type~I & $\Phi_2$ & $\Phi_2$ & $\Phi_2$ \\
type~II & $\Phi_2$ & $\Phi_1$ & $\Phi_1$ \\
type~III (lepton-specific) & $\Phi_2$ & $\Phi_2$ & $\Phi_1$ \\
type~IV (flipped) & $\Phi_2$ & $\Phi_1$ & $\Phi_2$ \\
\hline
\end{tabular}
\caption{Allowed fermion couplings in 
the four types of N2HDM.}
\label{tabtypes}
\end{center}
\end{table}
%%%%%%%%%%%%%%%%%%%%%%%%%%%%%%%%%%%%%%%%%%%%%%%%%%%%%%%%%%%%%%%%%%
Taking the electroweak symmetry breaking (EWSB) minima to be
charge and $\cp$-conserving, the scalar fields after EWSB
can be parametrised as
\begin{eqnarray}
\Phi_1 = \left( \begin{array}{c} \phi_1^+ \\ \frac{1}{\sqrt{2}} (v_1 +
    \rho_1 + i \eta_1) \end{array} \right) \;, \quad
\Phi_2 = \left( \begin{array}{c} \phi_2^+ \\ \frac{1}{\sqrt{2}} (v_2 +
    \rho_2 + i \eta_2) \end{array} \right) \;, \quad
\Phi_S = v_S + \rho_S \;, \label{eq:n2hdmvevs}
\end{eqnarray}
where $v_1, v_2, v_S$ are the real vevs acquired by the fields
$\Phi_1, \Phi_2$ and $\Phi_S$ respectively.
As in the 2HDM we define $\tb := v_2/v_1$.
As is evident from \refeq{eq:n2hdmvevs}, under such a field 
configuration, the $\cp$-odd and charged Higgs sector of the
N2HDM remain completely unaltered with respect to its
2HDM counterpart. However, the $\cp$-even scalar sector can 
undergo significant changes due the mixing among $\rho_1,
\rho_2$ and $\rho_S$, leading to a total of three 
$\cp$-even physical Higgses. Thus, a rotation from the
interaction to the physical basis can be achieved with the help of
a $3 \times 3$ orthogonal matrix as

\begin{eqnarray}
\left( \begin{array}{c} h_1 \\ h_2 \\ h_3 \end{array} \right) = R
\left( \begin{array}{c} \rho_1 \\ \rho_2 \\ \rho_S \end{array} \right).
\end{eqnarray}

We use the convention $m_{h_1} < m_{h_2} < m_{h_3}$ throughout the 
paper. The rotation matrix $R$ can be parametrized as
\begin{equation}
\label{mixingmatrix}
R=
\begin{pmatrix}
c_{\alpha_1}c_{\alpha_2} &
  s_{\alpha_1}c_{\alpha_2} &
    s_{\alpha_2} \\
-(c_{\alpha_1}s_{\alpha_2}s_{\alpha_3}+s_{\alpha_1}c_{\alpha_3}) &
  c_{\alpha_1}c_{\alpha_3}-s_{\alpha_1}s_{\alpha_2}s_{\alpha_3}  &
    c_{\alpha_2}s_{\alpha_3} \\
-c_{\alpha_1}s_{\alpha_2}c_{\alpha_3}+s_{\alpha_1}s_{\alpha_3} &
-(c_{\alpha_1}s_{\alpha_3}+s_{\alpha_1}s_{\alpha_2}c_{\alpha_3}) &
c_{\alpha_2}c_{\alpha_3}
\end{pmatrix}~,
\end{equation}
$\alpha_1, \alpha_2, \alpha_3$ being the three mixing angles, and
we use the short-hand notation $s_x = \sin x$, $c_x = \cos x$. The
singlet admixture of each physical state can be computed as 
$\Sigma_{h_i}= |R_{i3}|^2, i=1,2,3$.
The couplings of the Higgs bosons to SM particles are modified
w.r.t.\ the SM Higgs-coupling predictions due to the mixing in the Higgs
sector. It is convenient to express the couplings of the scalar
mass eigenstates $h_i$ normalized to the corresponding SM couplings.
We therefore introduce the coupling coefficients $c_{h_i V V}$ and
$c_{h_i f \bar f}$, such that the couplings to the massive vector bosons
are given by
\begin{equation}
\left(g_{h_i W W}\right)_{\mu\nu} =
\mathrm{i} g_{\mu\nu} \left(c_{h_i V V}\right) g M_W
\quad \text{and } \quad
\left(g_{h_i Z Z}\right)_{\mu\nu} =
\mathrm{i} g_{\mu\nu} \left(c_{h_i V V}\right) \frac{g M_Z}{\CW} \, ,
\end{equation}
where $g$ is the $SU(2)_L$ gauge coupling, $\CW$ the cosine of weak
mixing angle, $\CW = \MW/\MZ, \SW = \sqrt{1 - \CW^2}$,
and $M_W$ and $M_Z$ the masses of the $W$ boson
and the $Z$ boson, respectively. The couplings of the Higgs bosons
to the SM fermions are given by
\begin{equation}
g_{h_i f \bar{f}} =
\frac{m_f}{v} \left(c_{h_i f \bar{f}}\right) \; ,
\end{equation}
where $m_f$ is the mass of the fermion and
$v = \sqrt{(v_1^2 +v_2^2)}$ is the SM vev.
In \refta{tab:hvv}
we list the coupling coefficients for the couplings to
gauge bosons $V = W,Z$ for the three $\cp$-even Higgses.
They are identical in all four types of the (N)2HDM.
The same for the couplings to the fermions is listed in
\refta{tab:hff} for the four types of the N2HDM.
One can observe from \refta{tab:hff} that
the coupling pattern of the Yukawa sector in N2HDM is 
the same as that of 2HDM.

From \refeq{eq:scalarpot}, one can see that there are altogether
12 independent parameters in the model,
\begin{equation}
m_{11}^2 \; , \quad m_{22}^2 \;, \quad m_{12}^2 \;, \quad m_{S}^2 \;,
\quad \lambda_{i,~~i=1,8} \;.
\label{eq:original_inputs}
\end{equation}
However, one can use the 
three minimization conditions of the potential at the vacuum to substitute
the bilinears $m_{11}^2$, $m_{22}^2$ and $m_{S}^2$
for $v$, $\tan\beta$ and $v_S$.
Furthermore, the quartic couplings $\lambda_i$ can be replaced by the
physical scalar masses and mixing angles, leading to the
following parameter set~\cite{Muhlleitner:2016mzt};
\begin{equation}
\alpha_{1,2,3} \; , \quad \tan\beta \;, \quad v \; ,
\quad v_S \; , \quad m_{h_{1,2,3}} \;, \quad m_A \;, \quad \MHp
\;, \quad m_{12}^2 \; , \label{eq:inputs}
\end{equation}
where
$m_A$, $\MHp$ denote
the masses of the physical $\cp$-odd and charged Higgses respectively.
We use the code
\texttt{ScannerS}~\cite{Coimbra:2013qq,Muhlleitner:2016mzt} in our 
analysis to uniformly explore the set of independent parameters as
given in \refeq{eq:inputs} (see below).

%%%%%%%%%%%%%%%%%%%%%%%%%% TABLE %%%%%%%%%%%%%%%%%%%%%%%%%%%%%%%%
\begin{table}[t!]
% \begin{center}
\centering
\begin{tabular}{cc}
\hline
 &$c_{h_i VV} = c_\beta R_{i1} + s_\beta R_{i2}$ \\
\hline
$h_1$ & $c_{\alpha_2} c_{\beta-\alpha_1}$ \\
$h_2$ & $-c_{\beta-\alpha_1} s_{\alpha_2} s_{\alpha_3} + c_{\alpha_3}                                                                                 
s_{\beta-\alpha_1}$ \\
$h_3$ & $-c_{\alpha_3} c_{\beta-\alpha_1} s_{\alpha_2} - s_{\alpha_3}                                                                                 
s_{\beta-\alpha_1}$ \\
\hline
\end{tabular}
 \caption{The coupling factors of the neutral $\cp$-even Higgs bosons
   $h_i$ to the massive gauge bosons $V=W,Z$ in the N2HDM.} 
\label{tab:hvv}
% \end{center}
\end{table}
\noindent
%%%%%%%%%%%%%%%%%%%%%%%%%% TABLE %%%%%%%%%%%%%%%%%%%%%%%%%%%%%%%%
\begin{table}
\centering
% \begin{center}
\begin{tabular}{lccc} \hline
& $u$-type ($c_{h_i t\bar t}$)& $d$-type ($c_{h_i b\bar b}$)&
  leptons ($c_{h_i \tau\bar\tau}$)\\ \hline
type~I & $R_{i2} / s_\beta$
& $R_{i2} / s_\beta$ &
$R_{i2} / s_\beta$ \\
type~II & $R_{i2} / s_\beta $
& $R_{i1} / c_\beta $ &
$R_{i1} / c_\beta $ \\
type~III (lepton-specific) & $R_{i2} / s_\beta$
& $R_{i2} / s_\beta$ &
$R_{i1} / c_\beta$ \\
type~IV (flipped) & $R_{i2} / s_\beta$
& $R_{i1} / c_\beta$ &
$R_{i2} / s_\beta$ \\ \hline
\end{tabular}
\caption{Coupling factors of the Yukawa couplings of
   the N2HDM Higgs bosons $h_i$ w.r.t.\ their SM values.}
\label{tab:hff}
% \end{center}
\end{table}
%%%%%%%%%%%%%%%%%%%%%%%%%%%%%%%%%%%%%%%%%%%%%%%%%%%%%%%%%%%%%%%%%%%%%
In our analysis we will identify the lightest $\cp$-even Higgs boson,
$h_1$, with the one being potentially responsible for the signal at
$\sim 96 \gev$. The second lightest $\cp$-even Higgs boson will be
identified with the one observed at $\sim 125 \gev$.

%%%%%%%%%%%%%%%%%%%%%%%%%%%%%%%%%%%%%%%%%%%%%%%%%%%%%%%%%%%%%%%%%%%%%%%%%%
%%%%%%%%%%%%%%%%%%%%%%%%%%%%%%%%%%%%%%%%%%%%%%%%%%%%%%%%%%%%%%%%%%%%%%%%%%

\section {Relevant constraints}
\label{sec:constraints}

In this section we will describe the various theoretical and
experimental constraints considered in our scans.
The theoretical constraints are all implemented in \texttt{ScannerS}.
For more details, we refer the reader to the corresponding
references given 
below. The experimental constraints implemented in \texttt{ScannerS}
were supplemented with the most recent ones
by linking the parameter points from \texttt{ScannerS}
to the more recent versions of other public codes, which we will also
describe in more detail in the following. 

\subsection{Theoretical constraints}
\label{subsec:theo}
Like all models with extended scalar sectors, the N2HDM
also faces important constraints coming from tree-level
perturbartive unitarity, stability of the vacuum and the
condition that the vacuum should be a global minimum of
the potential. We briefly describe these constraints below.

\begin{itemize}
\item Tree-level perturbative unitarity conditions
ensure that the potential remains perturbative up to very high 
energy scales. This is achieved by demanding that the amplitudes of
the scalar quartic interactions leading to $2 \to 2$
scattering processes remain below the value of $8 \pi$ at tree-level.
The calculation was carried out in \citere{Muhlleitner:2016mzt} and
is implemented in \texttt{ScannerS}.

\item Boundedness from below demands 
that the potential remains positive when the field values approach infinity.
\texttt{ScannerS} automatically ensures that the N2HDM potential is bounded from
below by verifying that the necessary and sufficient conditions
as given in \citere{Klimenko:1984qx} are fulfilled.
The same conditions can be found in \citere{Muhlleitner:2016mzt}
in the notation adopted in this paper.

\item Following the procedure of \texttt{ScannerS}, we 
impose the condition that the vacuum should be
a global minimum of the potential. Although this condition 
can be avoided in the case of a metastable vacuum with the tunneling
time to the real minimum being larger than the age of the universe,
we do not explore this possibility in this analysis.
Details on the algorithm implemented in \texttt{ScannerS} to find
the global minimum of the potential can be found
in \citere{Coimbra:2013qq}. This algorithm has the advantage that
it works with the scalar masses and vevs as independent set of
parameters, which can be directly related to physical observables.
It also finds the global minimum of the potential without having
to solve coupled non-linear equations, therefore avoiding the usually numerically
most expensive task in solving the stationary conditions.

\end{itemize}

%%%%%%%%%%%%%%%%%%%%%%%%%%%%%%%%%%%%%%%%%%%%%%%%%%%%%%%%%%%%%%%%%%%%%%%%%%%%%%%

\subsection{Constraints from direct searches at colliders}
\label{subsec:collider}
Searches for charged Higgs bosons at the LHC are very effective
constraining the $\tb$-$\MHp$ plane of 2HDMs~\cite{Arbey:2017gmh}.
Since the charged scalar sector of the 2HDM is identical
to that of the N2HDM, the
bounds on the parameter space of the former also cover the 
corresponding parameter space of the latter. Important searches in our context are the direct
charged Higgs production $pp\to H^\pm t b$ with the decay modes
$H^\pm \to \tau \nu$ and $H^\pm \to tb$~\cite{Aaboud:2018cwk}.
The $95\%$ confidence level
exclusion limits of all important searches for charged Higgs bosons
are included in the public code
\texttt{HiggsBounds v.5.3.2}~\cite{Bechtle:2008jh,Bechtle:2011sb,Bechtle:2013gu,Bechtle:2013wla,Bechtle:2015pma}.
The theoretical cross section predictions for the production of the
charged Higgs at the LHC are provided by the LHC Higgs Cross Section Working
Group~\cite{Berger:2003sm,Flechl:2014wfa,Degrande:2015vpa,deFlorian:2016spz}.\footnote{We
thank T. Stefaniak for a program to extract the prediction
from the grid provided by
the LHC Higgs Cross Section Working Group.}
The rejected parameter points are concentrated in the region $\tb < 1$,
where the coupling of the charged Higgs to top quarks is
enhanced~\cite{Aaboud:2018cwk}. 
Bounds from searches for charged Higgs bosons at
LEP~\cite{Abbiendi:2013hk,Abdallah:2003wd,Abreu:1999bx,Achard:2003gt,Abbiendi:2008aa,Abbiendi:1998rd,Alexander:1995vg} are
irrelevant for our analysis, 
because constrains from flavor physics observables usually exclude very light
$H^\pm$ kinematically in the reach of LEP.

Direct searches for additional neutral Higgs bosons can exclude some of the
parameter points, mainly when the heavy Higgs boson $h_3$
or the $\cp$-odd Higgs boson $A$ are rather light.
\texttt{HiggsBounds} includes all relevant LHC searches for additional Higgs
bosons, such as possible decays of $h_3$ and $A$ to the singlet-like state
$h_1$ or the SM-like Higgs boson $h_2$.
The most relevant channels are the following:
CMS searched for pseudoscalars decaying into a $Z$-boson and scalar
in final states with two $b$-jets and two leptons, where the scalar
lies in the mass range of
$125\pm 10\gev$~\cite{CMS:2014yra,CMS:2018xvc}.
Both ATLAS and CMS searched for additional heavy Higgs bosons in the
$H\to ZZ$ decay channel including different final
states~\cite{Aad:2015kna,Aaboud:2017rel,CMS:2017vpy}.
For the flipped scenario, apart from the searches just mentioned,
also the search for $\cp$-even and -odd scalars decaying into a
$Z$-boson and a scalar, which then decays to a pair of
$\tau$-leptons~\cite{CMS:2015mba}, is relevant, because the coupling of
the light singlet-like scalar at $\sim 96\gev$ to $\tau$-leptons can be enhanced.
Of course, the light scalar is also directly constrained via
the Higgsstrahlung process with subsequent decay to a pair of
$b$-quarks at LEP~\cite{Schael:2006cr}
and by searches for diphoton resonances at the LHC
including all relevant production
mechanisms~\cite{ATLAS:2018xad,Sirunyan:2018aui}.
However, these constraints are weak, and they are the ones where
LEP and CMS saw the excesses we are investigating here.

%%%%%%%%%%%%%%%%%%%%%%%%%%%%%%%%%%%%%%%%%%%%%%%%%%%%%%%%%%%%%%%%%%%%%%%%%%%%%%%

\subsection{Constraints from the SM-like Higgs-boson properties}
\label{subsec:SMlike}
Any model beyond the SM has to accommodate the SM-like Higgs boson,
with mass and signal strengths as they were measured at the
LHC~\cite{Aad:2012tfa,Chatrchyan:2012xdj,Khachatryan:2016vau}.
In our scans the compatibility of the $\cp$-even scalar $h_2$ with a mass
of $125.09\gev$ with the measurements of signal strengths at Tevatron and LHC
is checked in a twofold way.

Firstly, the program \texttt{ScannerS}, that we
use to generate the benchmark points, contains an individual
check of the signal strengths
\begin{equation}
\frac{\mu_F}{\mu_V} \; , \quad
\mu_F^{\gamma\gamma} \; , \quad
\mu_F^{ZZ} \; , \quad
\mu_F^{WW} \; , \quad
\mu_F^{\tau\tau} \; , \quad
\mu_F^{bb} \; ,
\end{equation}
as they are quoted in \citere{Khachatryan:2016vau},
where an agreement within $\pm2 \, \sigma$ is required.
The signal strengths are defined as
\begin{equation}
\mu_F^{xx} = \mu_F \frac{\br_{\text{N2HDM}}(h_i \to xx)}
                        {\br_{\text{SM}}(H \to xx)} \; .
\end{equation}
Here, the production cross sections associated with couplings to
fermions, normalized to the SM prediction, 
are defined as
\begin{equation}
  \mu_F = \frac{\sigma_{\text{N2HDM}}(\rm ggF) + \sigma_{\text{N2HDM}}(bbH)}
               {\sigma_{\text{SM}}(\rm ggF)} \; ,
\end{equation}
where the production in association with a pair of $b$-quarks ($bbH$)
can be neglected in the SM,
whereas in N2HDM it has to be included since it can be enhanced by $\tan\beta$.
The cross section for vector boson fusion (VBF) production and the
associated production with a vector boson ($VH$) are given
by the coupling coefficient $c_{h_2 V V}$,
\begin{equation}
  \mu_V = \frac{\sigma_{\text{N2HDM}}({\rm VBF})}{\sigma_{\text{SM}}({\rm VBF})}
        = \frac{\sigma_{\text{N2HDM}}(VH)}{\sigma_{\text{SM}}(VH)}
        = c^2_{h_2 V V} \; ,
\end{equation}
where we made use of the fact that QCD corrections cancel in the ratio
of the vector boson fusion cross sections in the N2HDM and
the SM~\cite{Muhlleitner:2016mzt}. The ggF and $bbH$ cross sections are
provided by \texttt{ScannerS} with the help of data tables obtained using
the public code \texttt{SusHi}~\cite{Harlander:2012pb,Harlander:2016hcx}.
The couplings squared normalized to the SM prediction, for instance
$c^2_{h_i V V}$, are calculated via the interface of \texttt{ScannerS} with
the spectrum generator
\texttt{N2HDECAY}~\cite{Muhlleitner:2016mzt,Djouadi:1997yw,Butterworth:2010ym}.

In a second step, we supplemented the Higgs-boson data
from \citere{Khachatryan:2016vau} that is implemented
in \texttt{ScannerS} with the most recent Higgs-boson measurements: we 
verify the agreement of the generated points with all currently available
measurements using the public code
\texttt{HiggsSignals
v.2.2.3}~\cite{Bechtle:2013xfa,Stal:2013hwa,Bechtle:2014ewa}.
\texttt{HiggsSignals} provides a
statistical $\chi^2$ analysis of the SM-like Higgs-boson predictions of
a certain model compared to the measurement of Higgs-boson signal rates
and masses from Tevatron and LHC. The complete list of implemented
experimental data can be found in~\citere{higgssignals-www}.
In our scans, we will show the reduced $\chi^2$,
\begin{equation}
  \label{eq:chiHS}
  \chi_{\rm red}^2 = \frac{\chi^2}{n_{\rm obs}} \; ,
\end{equation}
where $\chi^2$ is provided by \texttt{HiggsSignals} and $n_{\rm obs}=101$ is
the number of experimental observations considered.

%%%%%%%%%%%%%%%%%%%%%%%%%%%%%%%%%%%%%%%%%%%%%%%%%%%%%%%%%%%%%%%%%%%%%%%%%%%%%%%

\subsection{Constraints from flavor physics}
\label{sub:flavor}

Constraints from flavor physics prove to be very significant
in the N2HDM because of the presence of the charged Higgs boson. Since
the charged Higgs sector of N2HDM is unaltered with respect to 
2HDM, we can translate the bounds from the 2HDM parameter space
directly on to our scenario for most of the observables. 
Various flavor observables like rare $B$~decays, 
$B$~meson mixing parameters, $\br(B \to X_s \gamma)$, 
LEP constraints on $Z$ decay partial widths
etc., which are sensitive to charged Higgs boson exchange, provide
effective constraints on the available 
parameter space~\cite{Enomoto:2015wbn,Arbey:2017gmh}. 
However, for
the low $\tb$ region that we are interested in (see below),
the constraints which must be taken into account 
are \cite{Arbey:2017gmh}: $\br(B \to X_s \gamma)$, 
constraints on $\Delta M_{B_s}$
from neutral $B-$meson mixing and $\br(B_s \to \mu^+ \mu^-)$. 
The dominant contributions
to the former two processes come from diagrams involving
$H^\pm$ and top quarks (see e.g.\ \citeres{Ciuchini:1997xe,Hermann:2012fc,Misiak:2015xwa} 
for $\br(B \to X_s \gamma)$
and \citeres{Buras:1989ui, Barger:1989fj, Chang:2015rva}
for $\Delta M_{B_s}$) and can be taken to be independent of the neutral
scalar sector to a very good approximation. Thus, the bounds
for them can be taken over directly from the 2HDM to our case.
Since the ${H^\pm t b}$ coupling  depends on the Yukawa sector 
of the model, the flavor bounds can differ for different N2HDM
types~\cite{Arbey:2017gmh}. Owing to identical quark Yukawa coupling
patterns, limits for type~I and~III scenarios come out to be very similar.
The same holds for type~II and~IV.
Constraints from $\br(B \to X_s \gamma)$ 
exclude $\MHp < 650\gev$ for all values of 
$\tan\beta \gtrsim 1$ in the type~II and~IV 2HDM, while for type~I 
and~III the bounds are more $\tan\beta-$dependent.
For $\MHp\geq 650\gev$ (as in our case) the dominant constraint
is the one obtained from the measurement of $\Delta M_{B_s}$.

For still lower values of $\tan\beta \lesssim 1$, bounds from
the measurement of $\br(B_s \to \mu^+ \mu^-)$ become
important~\cite{Arbey:2017gmh}.
Unlike the above two observables, $\br(B_s \to \mu^+ \mu^-)$
can get contributions from the neutral scalar
sector of the model as well~\cite{Li:2014fea,Cheng:2015yfu}. 
Thus, in principle 
the value of $\br(B_s \to \mu^+ \mu^-)$ in the
N2HDM may differ from that  of 2HDM because of additional contributions
coming from $h_1$ containing a large singlet component (see below).
However, we must note that the contributions from the N2HDM $\cp$-even
Higgs bosons can be expected to be small once we demand the presence of 
substantial singlet components in them, as it is the case
in our analysis. A detailed
calculation of various flavor observables in the specific case of the
N2HDM is beyond the scope of this work. Furthermore, as mentioned in
\refse{subsec:collider}, in the region $\tb \lesssim 1$, the constraints
from direct LHC searches of $H^\pm$ already provide fairly strong constraints.
Also the constraint from $\Delta M_{B_s}$ covers the region of very small
$\tb$.
Keeping the above facts in mind, in our work we use 
the flavor bounds for $\br(B \to X_s \gamma)$
and $\Delta M_{B_s}$ as obtained 
in \citere{Arbey:2017gmh} for different N2HDM types.

%%%%%%%%%%%%%%%%%%%%%%%%%%%%%%%%%%%%%%%%%%%%%%%%%%%%%%%%%%%%%%%%%%%%%%%%%%%%%%%

\subsection{Constraints from electroweak precision data}
\label{sec:stu}

Constraints from electroweak precision observables can in a simple
approximation be expressed in terms of the oblique parameters S, T and
U~\cite{Peskin:1990zt,Peskin:1991sw}. 
Deviations to these parameters are significant if new physics
beyond the SM enters mainly through gauge boson self-energies, as it is
the case for extended Higgs sectors.
\texttt{ScannerS} has implemented the one-loop corrections to the oblique
parameters for models with an arbitrary number of Higgs doublets
and singlets from \citere{Grimus:2008nb}. This calculation is valid under the
criteria that the gauge group is the SM $SU(2)\times U(1)$, and that
particles beyond the SM have suppressed couplings to light SM fermions.
Both conditions are fulfilled in the N2HDM.
Under these assumptions, the corrections are independent of the Yukawa
sector of the N2HDM, and therefore the same for all types.
The corrections to the oblique parameters are very sensitive to the relative
mass squared differences of the scalars. They become small when either
the heavy doublet-like Higgs $h_3$ or the $\cp$-odd scalar $A$ have a mass
close to the mass of the charged Higgs
boson~\cite{Bertolini:1985ia,Hollik:1986gg}.
In 2HDMs there is a strong correlation between $T$ and
$U$, and $T$ is the most sensitive of the three oblique parameters.
Thus, $U$ is much smaller in points not excluded by an extremely
large value of $T$~\cite{Funk:2011ad}, and the contributions
to $U$ can safely be dropped.
Therefore, for points to be in agreement
with the experimental observation, we
require that the prediction of the $S$ and the $T$ parameter
are within the $2 \, \sigma$ ellipse, corresponding to
$\chi^2=5.99$ for two degrees of freedom.

%%%%%%%%%%%%%%%%%%%%%%%%%%%%%%%%%%%%%%%%%%%%%%%%%%%%%%%%%%%%%%%%%%%%%%%%%%

\section{Experimental excesses}
\label{sec:excesses}

The main purpose of our analysis is to find a model that fits the two
experimental excesses in the Higgs boson searches at CMS and LEP.
As experimental input for the signal strengths we use the values
\begin{equation}
\mu_{\rm LEP} = 0.117 \pm 0.057 \quad \text{and} \quad
\mu_{\rm CMS} = 0.6   \pm 0.2   \; ,
\label{mumu}
\end{equation}
as quoted in \citeres{Schael:2006cr,Cao:2016uwt}
and~\cite{Sirunyan:2018aui,Shotkin:2017}. 

\smallskip
We evaluate the signal strengths for the excesses in the narrow width
approximation.
For the LEP excess this is given by, 
\begin{equation}
\label{eq:mulep}
\mu_{\rm LEP} =
  \frac{\sigma_{\rm N2HDM}(e^+e^-\to Z h_1)}
       {\sigma_{\SM}(e^+e^-\to Z H)}
  \cdot
  \frac{\br_{\rm N2HDM}(h_1\to b\bar{b})}
       {\br_{\SM}(H\to b\bar{b})} =
  \left|c_{h_1 V V}\right|^2
  \frac{\br_{\rm N2HDM}(h_1\to b\bar{b})}
       {\br_{\SM}(H \to b\bar{b})} \; ,
\end{equation}
where we assume that the cross section ratio can be
expressed via the effective coupling of $h_1$ to vector bosons normalized
to the SM prediction, which is provided by \texttt{N2HDECAY}, as
is the branching ratio of $h_1$ to two photons.
For the CMS signal strength one finds, 
\begin{equation}
\label{eq:mucms}
\mu_{\rm CMS} =
  \frac{\sigma_{\rm N2HDM}(gg \to h_1)}
       {\sigma_{\SM}(gg \to H))}
  \cdot
  \frac{\br_{\rm N2HDM}(h_1 \to \gamma\gamma)}
       {\br_{\SM}(H \to \gamma\gamma)} =
  \left|c_{h_1 t \bar{t}}\right|^2
  \frac{\br_{\rm N2HDM}(h_1 \to \gamma\gamma)}
       {\br_{\SM}(H \to \gamma\gamma)} \; .
\end{equation}
The SM predictions for the branching ratios and the cross section via ggF
can be found in \citere{Heinemeyer:2013tqa}. We checked that the approximation
of the cross section ratio in \refeq{eq:mucms} with $|c_{h_1 t \bar{t}}|^2$
is accurate to the percent-level by comparing
with the result for $\mu_{\rm CMS}$ evaluated with the
ggF cross section provided by \texttt{ScannerS}. Both approaches give
equivalent results considering the experimental uncertainty in $\mu_{\rm CMS}$.

%%%%%%%%%%%%%%%%%%%%% T A B L E %%%%%%%%%%%%%%%%%%%%%%%%%%%%%%%%%%%%%%%%%%%%%%
\begin{table}%[htb!]
\centering
% \begin{center}
\renewcommand{\arraystretch}{1.2}
\begin{tabular}{cccc} 
\hline
  & Decrease $c_{h_1 b \bar{b}}$ &
    No decrease $c_{h_1 t \bar{t}}$ &
    No enhancement $c_{h_1 \tau \bar{\tau}}$ \\
\hline
  type~I & \cmark $\;(\frac{R_{12}}{s_\beta})$ &
    \xmark $\;(\frac{R_{12}}{s_\beta})$ &
    \cmark  $\;(\frac{R_{12}}{s_\beta})$ \\
  type~II & \cmark  $\;(\frac{R_{11}}{c_\beta})$ &
    \cmark  $\;(\frac{R_{12}}{s_\beta})$ &
    \cmark  $\;(\frac{R_{11}}{c_\beta})$ \\
  lepton-specific & \cmark  $\;(\frac{R_{12}}{s_\beta})$ &
  \xmark  $\;(\frac{R_{12}}{s_\beta})$ &
  \xmark  $\;(\frac{R_{11}}{c_\beta})$ \\
  flipped & \cmark $\;(\frac{R_{11}}{c_\beta})$ &
  \cmark $\;(\frac{R_{12}}{s_\beta})$ &
  \xmark $\;(\frac{R_{12}}{s_\beta})$ \\
\hline
\end{tabular}
\caption{Conditions that have to be satisfied to accommodate the LEP and
CMS excesses simultaneously with a light $\cp$-even scalar $h_1$ with dominant
singlet component. In brackets we state the relevant coupling coefficients
$c_{h_1 f \bar{f}}$ for the conditions for each type.}
\label{tab:cond}
\renewcommand{\arraystretch}{1.0}
% \end{center}
\end{table}
%%%%%%%%%%%%%%%%%%%%% T A B L E %%%%%%%%%%%%%%%%%%%%%%%%%%%%%%%%%%%%%%%%%%%%%%

As can be seen from \refeqs{mumu} - (\ref{eq:mucms}), the CMS excess points
towards  the existence of a scalar with a SM-like production rate,
whereas the LEP excess demands that the scalar should have a 
squared coupling to massive vector bosons of $ \sim 0.1$ times that 
of the SM Higgs boson of the same
mass.
This suppression of the coupling coefficient $c_{h_1 V V}$
is naturally fulfilled for a singlet-like state, that acquires
its interaction to SM particles via a considerable mixing
with the SM-like Higgs boson, thus motivating the explanation
of the LEP excess with the real singlet of the N2HDM.
For the CMS excess, on the other hand, it appears to be difficult
at first sight
to accommodate the large signal strength, because one expects a
suppression of the loop-induced coupling to photons
of the same order as the one of $c_{h_1 V V}$, since in the SM the
Higgs-boson decay to photons is dominated by the $W$~boson loop.
However, it turns out that it is possible to overcompensate the
suppression of the loop-induced coupling to photons
by decreasing the
total width of the singlet-like scalar, leading to an
enhancement of the branching ratio of
the new scalar to the $\gamma \gamma$ final state.
In principle, the branching ratio
to diphotons can be further increased w.r.t.\ the SM by contributions
stemming from diagrams with the charged Higgs boson in the loop.
(In our scans, however,
these contributions are of minor significance due to the high
lower limit on the charged Higgs mass
of $650\gev$ from $\br(B_s \to X_s \gamma)$ constraints.)
The different types of N2HDM behave 
differently in this regard, based on how the doublet fields are
coupled to the quarks and leptons.
We summarize the general idea in \refta{tab:cond}
and argue that only the type~II and
type~IV (flipped) N2HDM can accommodate both excesses simultaneously
using a dominantly singlet-like scalar $h_1$ at $\sim 96\gev$.

The first condition is that the coupling of $h_1$ to $b$-quarks has to be
suppressed to enhance the decay rate to $\ga\ga$,
as the total decay width at this mass
is  still dominated by the decay to $b \bar b$.
At the same time one can not decrease the coupling to $t$-quarks too much,
because the decay width to photons strongly depends on the top quark 
loop contribution (interfering constructively with the charged Higgs
contribution). 
Moreover, the ggF production cross section is dominated at leading order
by the diagram with $t$-quarks in the loop. Thus, a decreased
coupling of $h_1$ to $t$-quarks implies a lower production cross section
at the LHC.
As one can deduce from \refta{tab:cond}, in type~I and the
lepton-specific N2HDM, the coupling coefficients
are the same for up- and down-type quarks. Thus, it is
impossible to satisfy both of the above criteria simultaneously 
in these models. Consequently, they fail to accommodate 
both the CMS and the LEP excesses.

One could of course go to the 2HDM-limit of the N2HDM by taking
the singlet scalar to be decoupled, and
reproduce the results observed previously in 
\citeres{Fox:2017uwr,Haisch:2017gql}, in which both excesses
are accommodated placing the second $\cp$-even Higgs boson
in the corresponding mass range. 
In the limit of the type~I 2HDM, the parameter space favorable
for the two excesses would correspond to very small values of coupling
of the $96 \gev$ state to up-type quarks,
because the dominant component of the scalar comes
from the down-type doublet field. This implies that
the ggF production mode no longer dominates the total
production cross section and the excesses can only be explained
by considering the contributions from other modes of
production like vector boson fusion and associated production
with vector bosons etc. The results for the lepton-specific 2HDM 
follow closely the ones for type~I because of similar
coupling structures in the two models. 
In  the CMS analysis \cite{Sirunyan:2018aui}, however,
the excess appears clearly in the ggF
production mode. Consequently, we discard these two versions of the N2HDM, as
they cannot provide a sufficiently large ggF cross section, while
yielding an adequate decay rate to $\ga\ga$ simultaneously.

Having discarded the type~I and type~III scenario, we now concentrate
on the remaining two possibilities.
In type~II and the flipped type~IV scenario, each of the doublet
fields $\Phi_1$ and $\Phi_2$
couple to either up- or down-type quarks, and it is possible to control
the size of the coupling coefficients
$c_{h_i t \bar{t}}$ and $c_{h_i b \bar{b}}$ independently.
Since the singlet-like scalar acquires its couplings to fermions
through the mixing with the doublet fields, this effectively leads
to one more degree of freedom to adjust its couplings independently
for up- and down-type quarks.
From the dependence of the mixing matrix elements $R_{11}$ and $R_{12}$
on the mixings angles $\alpha_i$, as stated in \refeq{mixingmatrix},
one can deduce that the relevant parameter in this case is $\alpha_1$.
For $|\alpha_1| \to \pi / 2$ the up-type doublet component of $h_1$
is large and the down-type doublet component goes to zero.
Thus, large values of $\alpha_1$ will correspond to an enhancement of
the branching ratio to photons,
because the dominant decay width to $b$-quarks, and therefore
the total width of $h_1$, is suppressed.

A third condition, although not as significant as the other two, is
related to the coupling of $h_1$ to leptons. If it
is increased, the decay to a pair of $\tau$-leptons will be
enhanced. Similar to the decay to $b$-quarks, it will compete with
the diphoton decay and can suppress the signal strength needed for
the CMS excess. The $\tau$-Yukawa coupling is not as
large as the $b$-Yukawa coupling, so this condition is not as important
as the other two. Still, as we will see in our numerical evaluation,
it is the reason why it is easier to fit the CMS excess in the type~II model
compared to the flipped scenario.
In the latter the coupling coefficient to leptons is equal to the one to
up-type quarks. 
Thus, in the region where the diphoton decay width is large, also decay
width to $\tau$-pairs is large, and both channels will compete.
In the type~II scenario, on the other hand, the coupling to leptons is
equal to the coupling to down-type quarks, meaning that in the relevant
parameter region both the decay to $b$-quarks and the decay to $\tau$-leptons
are suppressed.

In our scans we indicate the ``best-fit point'' referring to the
point with the smallest $\chi^2$ defined by
\begin{equation}
  \label{eq:chilepcms}
  \chi_{\rm CMS-LEP}^2 = \frac{(\mu_{\mathrm{LEP}} - 0.117)^2}{0.057^2} +
           \frac{(\mu_{\mathrm{CMS}} -   0.6)^2}{0.2^2   } \; ,
\end{equation}
quantifying the quadratic deviation w.r.t.\ the measured values, assuming
that there is no correlation between the signal strengths of
the two excesses. In principle, we could have combined the $\chi^2$ obtained
from \texttt{HiggsSignals} regarding the SM-like Higgs boson observables
with the $\chi^2$ defined above regarding the LEP and the CMS excesses.
In that case, however, the total $\chi^2$
would be strongly dominated by the SM-like Higgs boson contribution
from \texttt{HiggsSignals} due to the sheer amount of
signal strength observables implemented. Consequently, we refrain from
performing such a combined $\chi^2$ analysis.

%%%%%%%%%%%%%%%%%%%%%%%%%%%%%%%%%%%%%%%%%%%%%%%%%%%%%%%%%%%%%%%%%%%%%%%%%%%%%%%

%\newpage

\subsection*{ATLAS limits}

ATLAS pubished first Run\,II~results in the~$\ga\ga$~searches
below~$125$\,GeV with~$80$\,fb$^{-1}$~\cite{ATLAS:2018xad}.
No significant excess above
the~SM~expectation was observed in the mass range between $65$~and
$110 \gev$. 
However, the limit on cross section times branching ratio obtained in the
diphoton final state by ATLAS is substantially weaker than the
corresponding upper limit obtained by CMS at $\sim 96 \gev$. It does not
touch the $1\,\sig$ ranges of $\mu_{\rm CMS}$. Interestingly, the ATLAS
result shows a little ``shoulder'' (upward ``bump'') around $96 \gev$.
This was illustrated and discussed in Fig.~1 of \citere{Heinemeyer:2018wzl}.

%%%%%%%%%%%%%%%%%%%%%%%%%%%%%%%%%%%%%%%%%%%%%%%%%%%%%%%%%%%%%%%%%%%%%%%%%%
%%%%%%%%%%%%%%%%%%%%%%%%%%%%%%%%%%%%%%%%%%%%%%%%%%%%%%%%%%%%%%%%%%%%%%%%%%

\section{Results}
\label{sec:results}

In the following we will present our analyses in the type~II and
type~IV scenario.
The scalar mass eigenstate with dominant singlet-component will
be responsible for accommodating the LEP and the CMS excesses at
$\sim95$-$98\gev$.
As already mentioned above, we performed a
scan over the relevant parameters using the public code \texttt{ScannerS}.
We give the ranges of the free parameters for each type in the corresponding
subsection. We will make use of the possibility to set additional constraints
on the singlet admixture of each $\cp$-even scalar particle, which is
provided by \texttt{ScannerS}. Additional constraints on the mixing angles
$\alpha_i$, as will be explained in the following,
were implemented by us within the
appropriate routines to exclude irrelevant parameter space.

In our plots we will show the benchmark points that pass all the theoretical
and experimental constraints described in \refse{sec:constraints},
if not said otherwise.
We will provide details on the best-fit points for both types of
the N2HDM and explain
relevant differences regarding the contributions to the excesses at
LEP and CMS. 

Similar scans have been performed also for the N2HDM type~I
and~III (lepton specific).
We confirmed the negative result expected from the arguments
given in \refse{sec:excesses}. Consequently, we do not show any of these
results here, but concentrate on the two models that indeed can describe
the CMS and LEP excesses.

To enforce that the lightest scalar has the dominant singlet admixture,
we impose
\begin{equation}
65\% \leq \Sigma_{h_1} \leq 90\% \; ,
\end{equation}
while for the SM-like Higgs boson we impose a lower limit on the
singlet admixture of
\begin{equation}
\Sigma_{h_2} \geq 10\% \; .
\end{equation}
This assures that there is at least some doublet component in $h_1$ in each
scan point, which is necessary to fit the excesses.
We have checked explicitely that this bound has no impact on the
parameter space that have $\chi^2_{\rm CMS-LEP} \le 2.30$ (i.e.\ the
$1\,\sig$ range, see the discussion of \reffi{fig:ty2ty4sH1} below).
The conditions on the singlet admixture of the mass eigenstates can
trivially be translated into bounds on the mixing angles $\alpha_i$.
Taking into account that we want to increase the up-type component
of $h_1$ compared to its down-type component, one can deduce from
the definition of the mixing matrix in \refeq{mixingmatrix} that
$\alpha_1 \to \pm \pi/2$ is a necessary condition.
In this limit the coupling
coefficients of the SM-like Higgs boson $h_2$ to quarks can be
approximated by
\begin{equation}
c_{h_2 t \bar t} \sim \mp \frac{s_{\alpha_2} s_{\alpha_3}}{s_\beta} \quad
\text{and} \quad
c_{h_2 b \bar b} \sim \mp \frac{c_{\alpha_3}}{c_\beta} \; .
\end{equation}
Consequently, if $\alpha_2$ and $\alpha_3$ would have opposite signs,
one would be in the wrong-sign Yukawa coupling regime.
In this regime it is harder to accommodate the SM-like
Higgs boson properties, especially for low values of $\tan\beta$.
Also the possible singlet-component
of $h_2$ is more limited~\cite{Muhlleitner:2016mzt}.
To avoid entering the wrong-sign Yukawa coupling regime, we therefore impose
additionally
\begin{equation}
\alpha_2 \cdot \alpha_3 > 0 \; .
\label{eq:alal}
\end{equation}

%%%%%%%%%%%%%%%%%%%%%%%%%%%%%%%%%%%%%%%%%%%%%%%%%%%%%%%%%%%%%%%%%%%%%%%%%%

\subsection{Type~II}
\label{sec:type2}

Following the discussion about the various experimental and theoretical
constrains we chose to scan the following range of input parameters:
\begin{align}
95 \gev \leq m_{h_1} \leq 98 \gev \; ,
\quad m_{h_2} = 125.09 \gev \; ,
\quad 400 \gev \leq m_{h_3} \leq 1000 \gev \; , \notag \\
400 \gev \leq m_A \leq 1000 \gev \; ,
\quad 650 \gev \leq \MHp \leq 1000 \gev \; , \notag\\
0.5 \leq \tb \leq 4 \; ,
\quad 0 \leq m_{12}^2 \leq 10^6 \gev^2 \; ,
\quad 100 \gev \leq v_S \leq 1500 \gev \; . \label{eq:ranges}
\end{align}

%%%%%%%%%%%%%%%%%%%%%%%%%%%% F I G U R E %%%%%%%%%%%%%%%%%%%%%%%%%%%%%%
\begin{figure}%[htb!]
  \centering
  \includegraphics[width=0.8\textwidth]{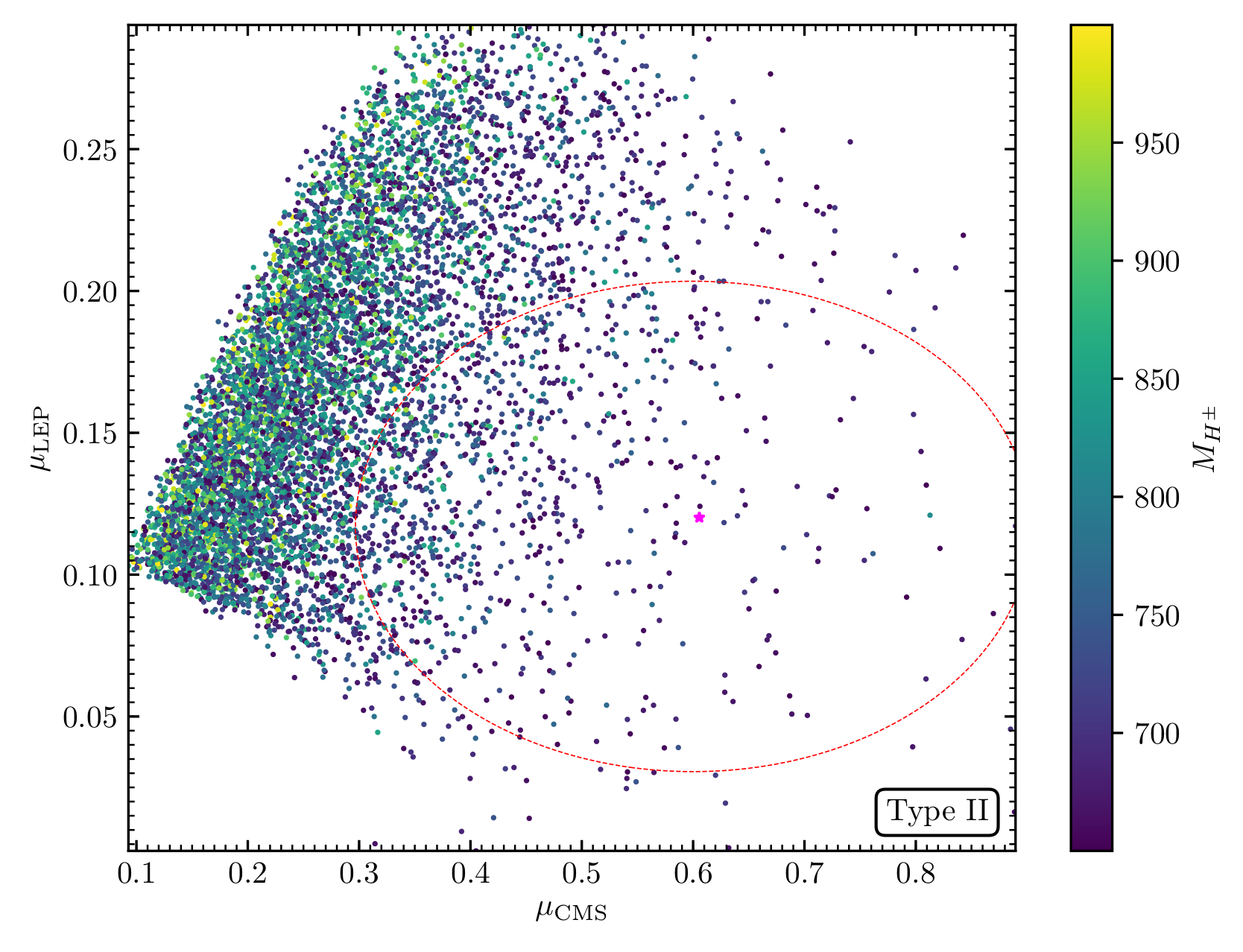}
  \vspace{-0.5cm}
  \caption{Type~II: the signal strengths $\mu_{\rm CMS}$ and
  $\mu_{\rm LEP}$ are shown 
  for each scan point respecting the experimental and
  theoretical constrains. The $1 \, \sigma$-region of both excesses
  is shown by the red ellipse. The colors show the
  mass of the charged Higgs. The magenta star is the best-fit point.
  The lowest (highest) value of $\MHp$ inside the
  $1 \, \sigma$ ellipse   is $650.03 \; (964.71) \gev$.}
  \label{fig:2Hpplot}
\end{figure}
%%%%%%%%%%%%%%%%%%%%%%%%%%%% F I G U R E %%%%%%%%%%%%%%%%%%%%%%%%%%%%%%

%%%%%%%%%%%%%%%%%%%%%%%%%%%% F I G U R E %%%%%%%%%%%%%%%%%%%%%%%%%%%%%%
\begin{figure}%[htb!]
  \centering
  \includegraphics[width=0.8\textwidth]{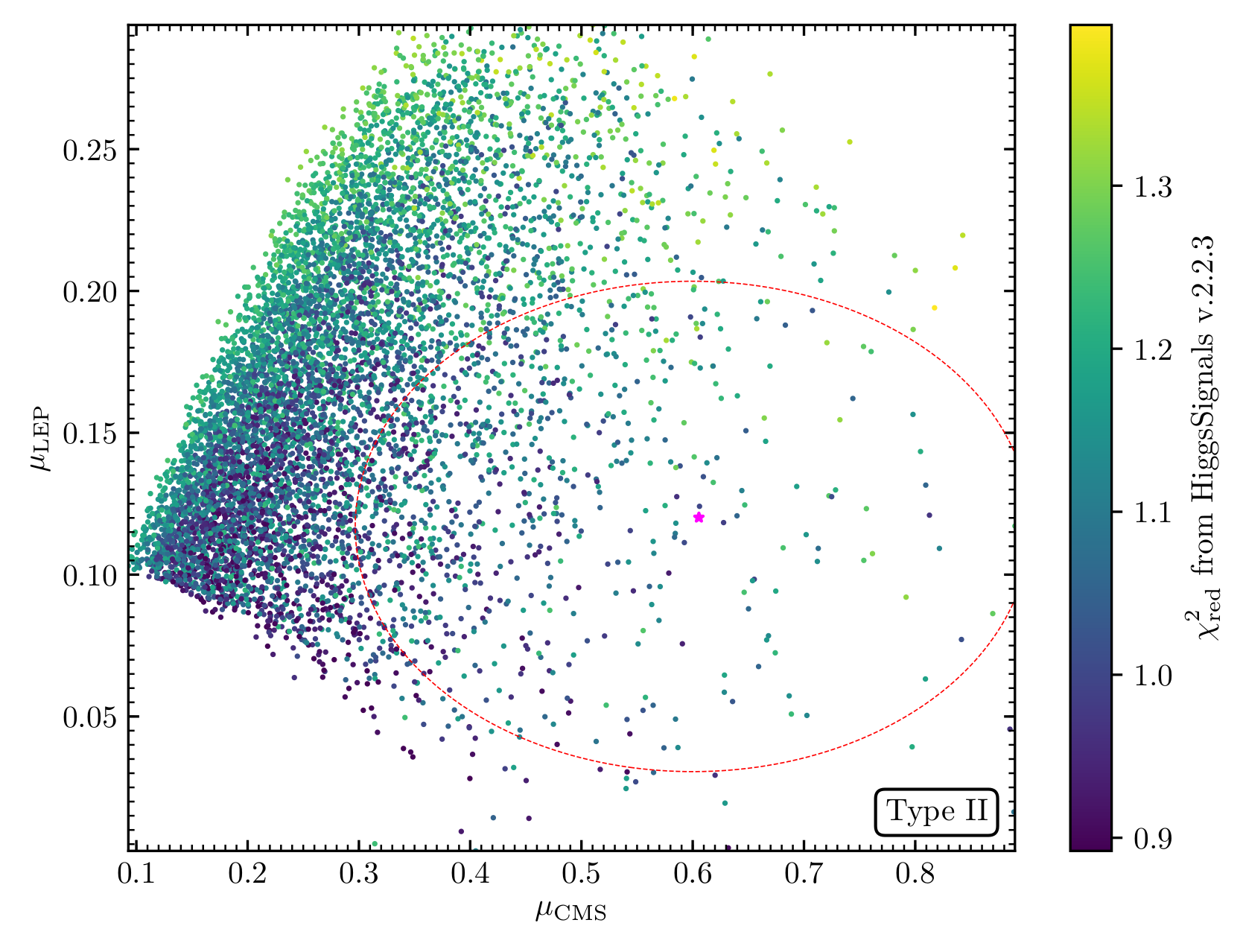}
%  \vspace{-0.5cm}
  \caption{Type~II: as in \reffi{fig:2Hpplot}, but here the colors indicate
  the $\chi_{\rm red}^2$ from HiggsSignals. The best-fit point (magenta)
  has $\chi_{\rm red}^2=1.237$ with 101 observations considered.
  The lowest (highest) value of $\chi_{\rm red}^2$ inside the
  $1 \, \sigma$ ellipse is $0.9052$ ($1.3304$).}
  \label{fig:2HSplot}
\end{figure}
%%%%%%%%%%%%%%%%%%%%%%%%%%%% F I G U R E %%%%%%%%%%%%%%%%%%%%%%%%%%%%%%

%%%%%%%%%%%%%%%%%%%%%%%%%%%% F I G U R E %%%%%%%%%%%%%%%%%%%%%%%%%%%%%%
\begin{figure}%[htb!]
  \centering
  \includegraphics[width=0.8\textwidth]{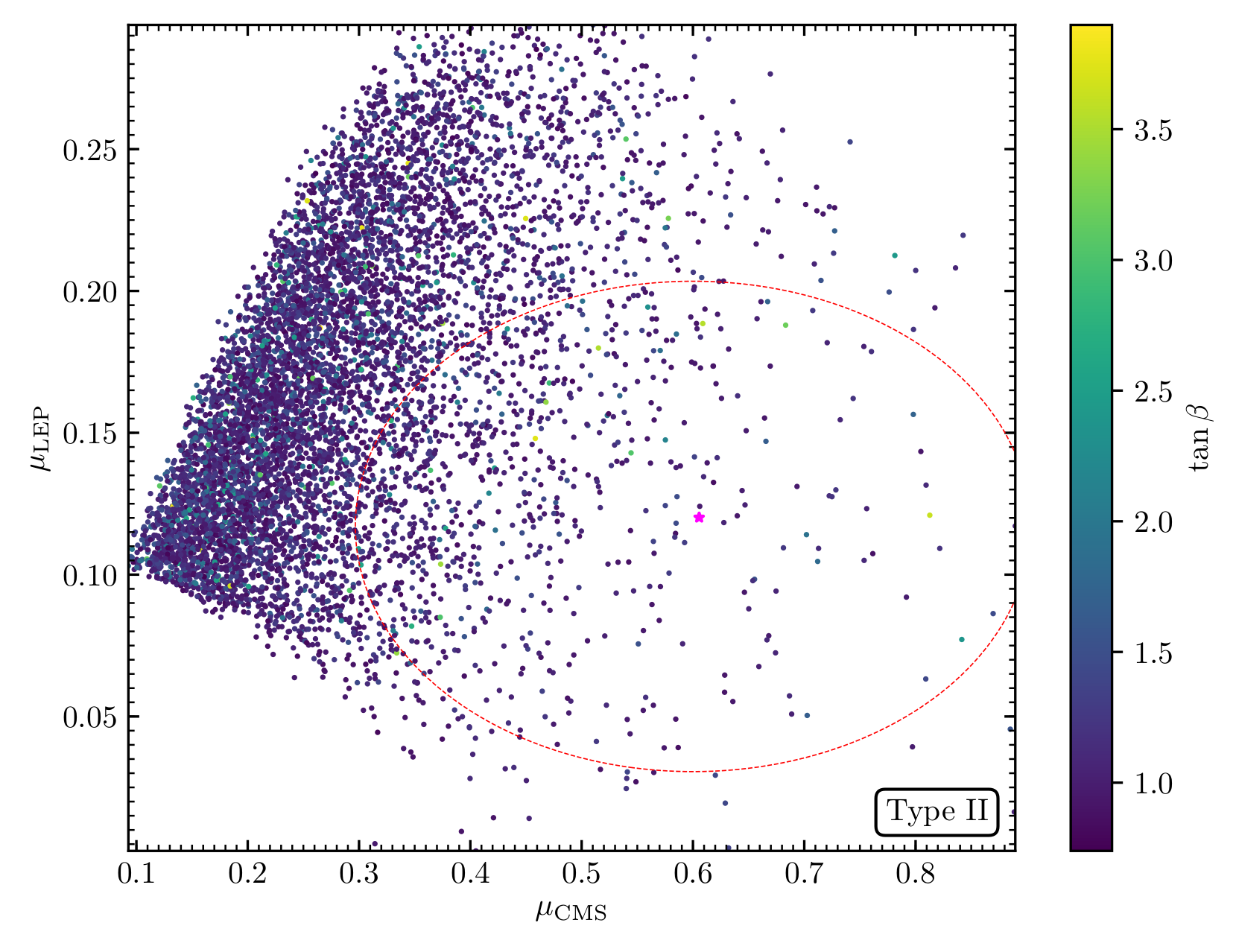}
%  \vspace{-0.5cm}
  \caption{Type~II: as in \reffi{fig:2Hpplot}, but here the colors indicate
  the value of $\tan\beta$. The lowest (highest) value of
  $\tan\beta$ inside the $1 \, \sigma$ ellipse is $0.7970$ ($3.748$).}
  \label{fig:2TBplot}
\end{figure}
%%%%%%%%%%%%%%%%%%%%%%%%%%%% F I G U R E %%%%%%%%%%%%%%%%%%%%%%%%%%%%%%
%%%%%%%%%%%%%%%%%%%%%%%%%%%% F I G U R E %%%%%%%%%%%%%%%%%%%%%%%%%%%%%%
\begin{figure}%[htb!]
\vspace{2em}
  \centering
  \begin{subfigure}[b]{0.48\linewidth}
    \centering\includegraphics[width=\textwidth]{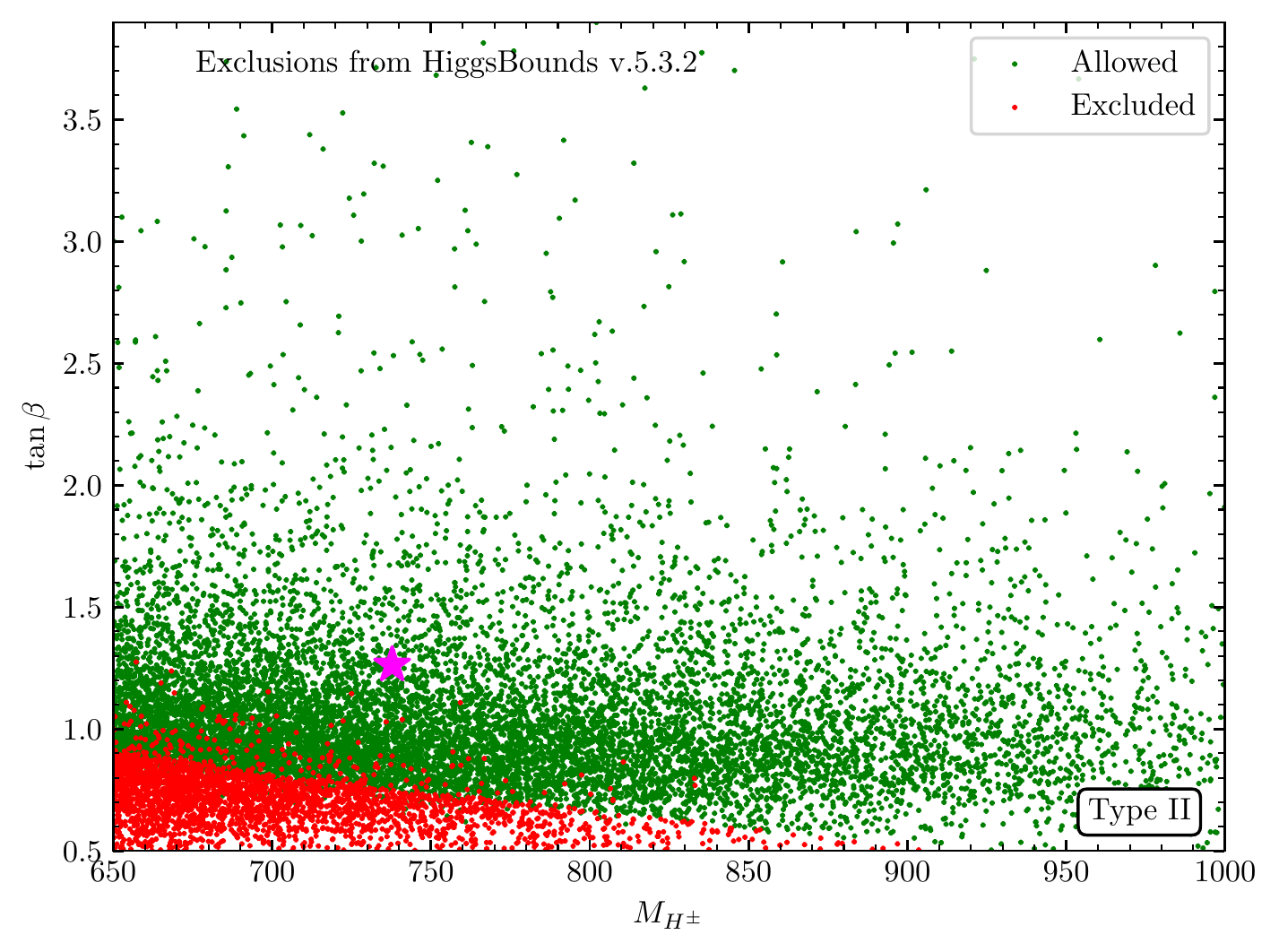}
    \caption{Direct searches at colliders in type~II.}
    \label{fig:2HBplot}
  \end{subfigure}
  ~
  \begin{subfigure}[b]{0.48\linewidth}
    \centering\includegraphics[width=\textwidth]{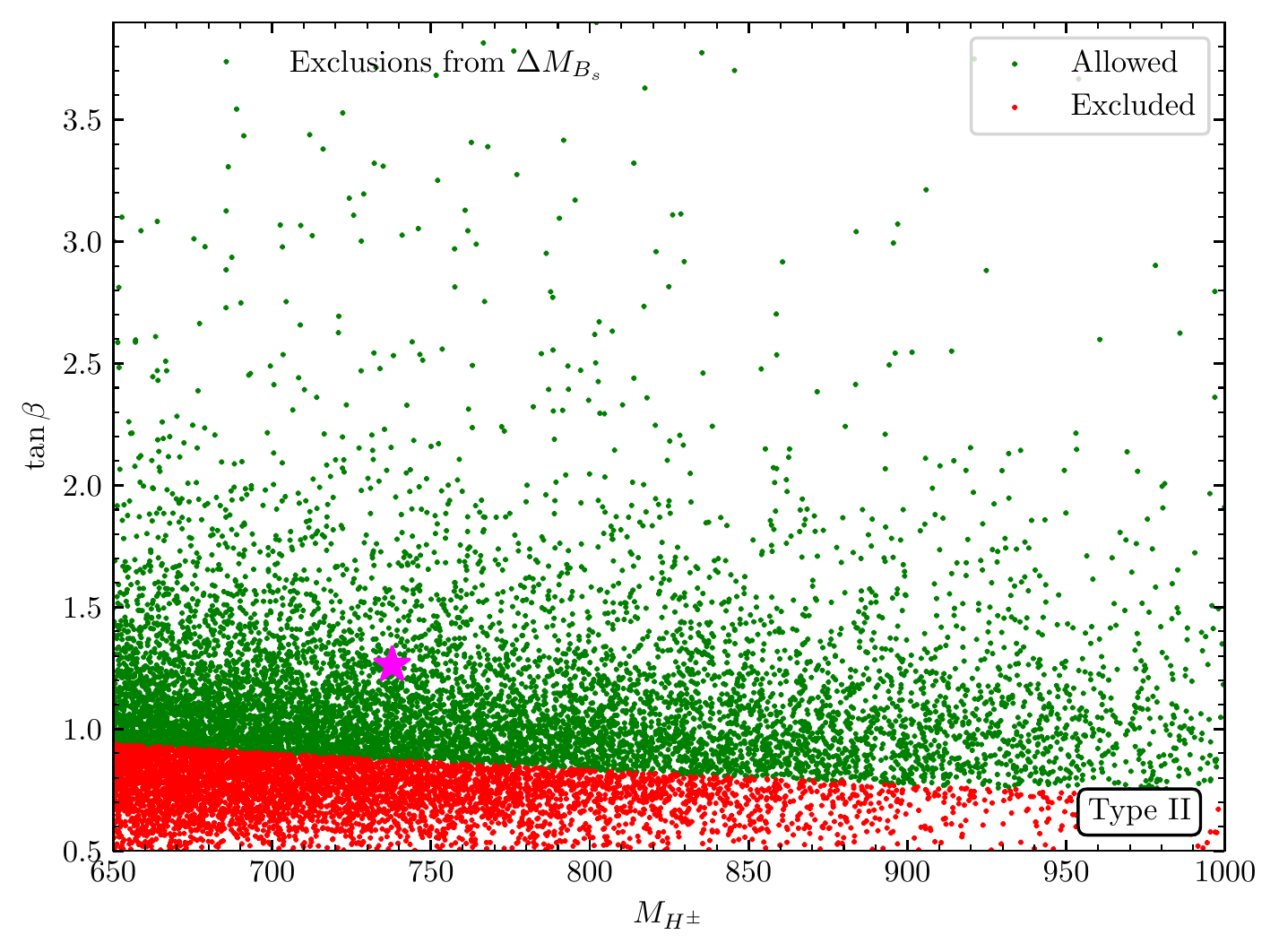}
    \caption{Flavor physics in type~II.}
    \label{fig:2FLplot}
  \end{subfigure}
  \caption{Allowed (\textit{green}) and excluded 
  (\textit{red}) points 
  considering direct searches (\textit{left})
  and flavor physics (\textit{right}) in the $\MHp$-$\tb$ plane.
  The magenta star is the best-fit point.}
\end{figure}
%%%%%%%%%%%%%%%%%%%%%%%%%%%% F I G U R E %%%%%%%%%%%%%%%%%%%%%%%%%%%%%%

We show the result of our scan in \reffis{fig:2Hpplot}-\ref{fig:2TBplot}
in the plane of the signal strengths $\mu_{\rm LEP}$ and $\mu_{\rm CMS}$ for each
scan point, where the best-fit point w.r.t.\ the two excesses is
marked by a magenta star. It should be kept in mind that the density of
points has no physical meaning and is a pure artifact of the ``flat
prior'' in our parameter scan. 
The red dashed line corresponds to the $1\,\sig$ ellipse, i.e.,
to $\chi_{\rm CMS-LEP}^2 = 2.30$ for two degrees of freedom,
with $\chi_{\rm CMS-LEP}^2$ defined in \refeq{eq:chilepcms}.
The colors of the points indicate the value
of the charged Higgs mass in \reffi{fig:2Hpplot} and the reduced
$\chi^2$ (see \refeq{eq:chiHS})
from the test of the SM-like Higgs-boson properties
with \texttt{HiggsSignals} in \reffi{fig:2HSplot}. 
One sees that various points fit both excesses simultaneously while also
accommodating the properties of the SM-like Higgs boson at $125\gev$.
From \reffi{fig:2Hpplot} we can conclude that lower values
for $\MHp$ are preferred to fit the
diphoton excess.
We emphasize that the dependence of the branching ratio of
$h_1$ to diphotons, and therefore of $\mu_{\mathrm{CMS}}$,
on $\MHp$ is due to the positive correlation between $\MHp$
and the total decay width of $h_1$.
The additional contributions
to the diphoton decay width of diagrams with the charged Higgs boson
in the loop has a minor dependence on $\MHp$ for $\MHp > 650\gev$.
When $\MHp$ becomes larger the constraints from the oblique parameters induce
also larger masses of the heavy Higgs $m_{h_3}$ and
the $\cp$-odd Higgs $m_A$.
These masses are correlated to the mixing angles in the scalar
sector via the tree-level perturbative unitarity and global minimum conditions.
Concerning a large suppression of the total decay width of $h_1$,
and thus an enhancement of $\br^{\gamma\gamma}_{h_1}$,
it turns out to be more difficult to achieve for larger $m_{h_3}$,
$\MHp$ and $M_A$.
Finally, we show in \reffi{fig:2TBplot} a plot with the colors indicating
the value of $\tan\beta$ in each point. An overall tendency
can be observed
that values of about $\tan\beta\sim 1$ are preferred in our scan.
However, we find points covering the whole $\tan\beta$-range used in our scan
within the $1 \, \sigma$ ellipse of the excesses.

The preferred low values of the charged Higgs mass and $\tan\beta$
give rise to the fact that the scenario presented here
will be in reach of direct searches for charged Higgs bosons at the
LHC~\cite{Aaboud:2018cwk}
(see the discussion in \refse{sec:propsindirect}). 
Already now, parts of the parameter space scanned here are excluded by
direct searches. This is illustrated in \reffi{fig:2HBplot}, where we
show points allowed by \texttt{HiggsBounds} in green, and the
excluded points in red. For values
of $\tb < 1$ direct searches are very constraining. The
experimental analysis responsible for this excluded region is the search for
charged Higgs bosons produced in association with a $t$- and a $b$-quark, and
the subsequent decay of the charged Higgs boson to a $tb$-pair,
performed by ATLAS~\cite{Aaboud:2018cwk}.
Apart from that, flavor physics can provide very strict bounds in the
$\MHp$-$\tb$ plane
(see the discussion in \refse{sub:flavor}).
We show the excluded
regions in our scan in \reffi{fig:2FLplot}.
We see that in the region of
lower values of the charged Higgs-boson mass, where the excesses are
reproduced most ``easily'', bounds from flavor physics are as good as 
the direct searches for additional Higgs bosons in the low $\tb$ region. 
Values of $\tb < 0.7$ are ruled out for the whole range of $\MHp$.

In \refta{tab:2best} we show the values of the free parameters
and the relevant branching ratios of the singlet-like scaler $h_1$, 
the SM-like Higgs boson $h_2$ as well as all other (heavier)
Higgs bosons of the model for the best-fit point of our scan,
which is highlighted with a magenta star
in \reffis{fig:2Hpplot}-\ref{fig:2FLplot}.
Remarkably, the branching ratio for the singlet-like scalar
to photons is larger than the one of the SM-like Higgs boson. As explained
in the beginning of \refse{sec:results} this is achieved by a
value of $\alpha_1 \sim \pi / 2$, 
which suppresses the decay to $b$-quarks and $\tau$-leptons, without
decreasing the coupling to $t$-quarks.
The most important BRs for the heavy Higgs bosons are those to the
heaviest quarks, $h_3 \to t \bar t$, $A \to t \bar t$ and $H^\pm \to tb$,
offering interesting prospects for future searches, as will be briefly
discussed in \refse{sec:future}.
Constraints from the oblique parameters
lead to a $\cp$-odd Higgs boson mass $m_A$ close to the mass of the charged
Higgs boson.
We stress, however, that this is not the only possibility to fulfill
the constraints from the oblique parameters. The alternative
possibility that $m_{h_3} \sim \MHp$ occurs as often as
$m_{A} \sim \MHp$ in our scan.
The value of $\tb$ is close to one, meaning that the benchmark point
shown here might be in range of future improved constraints
both from direct searches at colliders as well as from flavor physics.
More optimistically speaking, deviations from the SM predictions are expected
in those observables if our explanation of the LEP and CMS excesses
are implemented by nature.
We will discuss in \refse{sec:future} the prospects of detecting
deviations from the SM-prediction, that accompany our explanation of the
LEP and the CMS excess, at future colliders.

%%%%%%%%%%%%%%%%%%%%%%%%%% T A B L E %%%%%%%%%%%%%%%%%%%%%%%%%%%%%%%%%%%%%%%%%%
\begin{table}
\centering
\renewcommand{\arraystretch}{1.2}
\begin{tabular}{c c c c c c c}
 $m_{h_1}$ & $m_{h_2}$ & $m_{h_3}$ & $m_A$ & $\MHp$ & & \\
 \hline
 $96.5263$ & $125.09$ & $535.86$ & $712.578$ & $737.829$ & & \\
 \hline
 \hline
 $\tb$ & $\alpha_1$ & $\alpha_2$ & $\alpha_3$ & $m_{12}^2$ & $v_S$ & \\
 \hline
 $1.26287$ & $1.26878$ & $-1.08484$ & $-1.24108$ &
   $80644.3$ & $272.72$ & \\
  \hline
  \hline
 $\br^{bb}_{h_1}$ & $\br^{gg}_{h_1}$ & $\br^{cc}_{h_1}$ &
   $\br^{\tau\tau}_{h_1}$ & $\br^{\gamma\gamma}_{h_1}$ 
 & $\br^{WW}_{h_1}$  & $\br^{ZZ}_{h_1}$ \\
 \hline
 $0.5048$ & $0.2682$ & $0.1577$ & $0.0509$ 
    & $2.582\cdot 10^{-3}$ & $0.0137$ & $1.753\cdot 10^{-3}$ \\
 \hline
 \hline
 $\br^{bb}_{h_2}$ & $\br^{gg}_{h_2}$ & $\br^{cc}_{h_2}$ &
   $\br^{\tau\tau}_{h_2}$ & $\br^{\ga\ga}_{h_2}$ &
     $\br^{WW}_{h_2}$ & $\br^{ZZ}_{h_2}$ \\
 \hline
 $0.5916$ & $0.0771$ & $0.0288$ & $0.0636$ &
   $2.153\cdot 10^{-3}$ & $0.2087$ & $0.0261$ \\
 \hline
 \hline
 $\br^{tt}_{h_3}$ & $\br^{gg}_{h_3}$ & $\br^{h_1 h_1}_{h_3}$ &
   $\br^{h_1 h_2}_{h_3}$ & $\br^{h_2 h_2}_{h_3}$ &
     $\br^{WW}_{h_3}$ & $\br^{ZZ}_{h_3}$ \\
 \hline
 $0.8788$ & $2.537\cdot 10^{-3}$ & $0.0241$ &
   $0.0510$ & $3.181\cdot 10^{-3}$ &
     $0.0261$ & $0.0125$ \\
 \hline
 \hline
 $\br^{tt}_{A}$ & $\br^{gg}_{A}$ & $\br^{Z h_1}_{A}$ &
   $\br^{Z h_3}_{A}$ & $\br^{bb}_{A}$ & & \\
 \hline
 $0.6987$ & $1.771\cdot 10^{-3}$ & $0.1008$ &
   $0.1981$ & $5.36\cdot 10^{-4}$ & $ $ \\
 \hline
 \hline
 $\br^{tb}_{\Hpm}$ & $\br^{W h_3}_{\Hpm}$ & $\br^{W h_1}_{\Hpm}$ & & & \\
 \hline
 $0.6000$ & $0.3004$ & $0.0984$ & & &
\end{tabular}
\caption{Parameters of the best-fit point and branching ratios of the
scalars in the type~II scenario.
Dimensionful parameters are given
in GeV and the angles are given in radian.}
\label{tab:2best}
\renewcommand{\arraystretch}{1.0}
\end{table}
%%%%%%%%%%%%%%%%%%%%%%%%%% T A B L E %%%%%%%%%%%%%%%%%%%%%%%%%%%%%%%%%%%%%%%%%%

%%%%%%%%%%%%%%%%%%%%%%%%%%%%%%%%%%%%%%%%%%%%%%%%%%%%%%%%%%%%%%%%%%%%%%%%%%

\subsection{Type~IV (flipped)}
\label{sec:type4}

In the type~IV (flipped) scenario the couplings of the scalars to quarks
are unchanged with respect to the type~II scenario.
The coupling to leptons, however, is equal to the coupling to
the up-type quarks, instead of being equal to the coupling to down-type quarks,
as it is in the type~II scenario.
This means that while the parameter space that can fit the LEP and the CMS
excesses will be very similar to the one in the type~II analysis,
the non-suppression of the decay width of $h_1$ to $\tau$-leptons will
have to be compensated. Apart from that,
constrains especially from the SM-like Higgs boson measurements
and from direct searches will be different
(see also \refse{sec:future}).
For the scan in the type~IV scenario we choose the same range of parameters
as in the type~II scenario, shown in \refeq{eq:ranges}.
As explained in the beginning of \refse{sec:results} we further
impose \refeq{eq:alal}. 

%%%%%%%%%%%%%%%%%%%%%%%%%%%% F I G U R E %%%%%%%%%%%%%%%%%%%%%%%%%%%%%%
\begin{figure}%[htb!]
  \centering
  \includegraphics[width=0.8\textwidth]{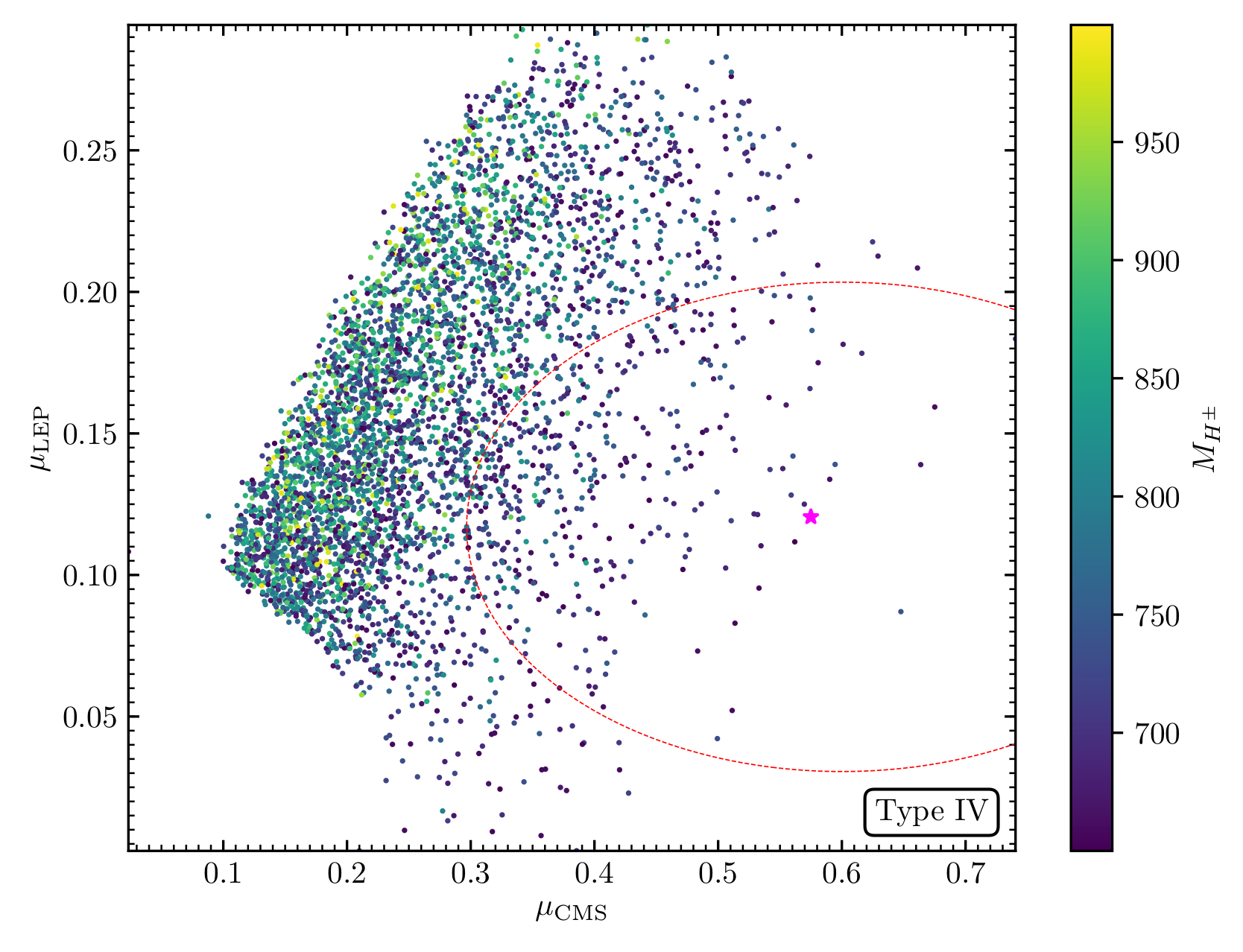}
  \vspace{-0.5cm}
  \caption{Type~IV: the signal strengths $\mu_{\rm CMS}$ and $\mu_{\rm LEP}$
  for each scan point respecting the experimental and
  theoretical constrains. The $1 \, \sigma$-region of both excesses
  is shown by the red ellipse. The colors show the
  mass of the charged Higgs. The magenta star indicates the best-fit point.
  The lowest (highest) value of $\MHp$ inside the
  $1 \, \sigma$ ellipse is $650.01 \; (931.85)\gev$.}
  \label{fig:4Hpplot}
\end{figure}
%%%%%%%%%%%%%%%%%%%%%%%%%%%% F I G U R E %%%%%%%%%%%%%%%%%%%%%%%%%%%%%%

%%%%%%%%%%%%%%%%%%%%%%%%%%%% F I G U R E %%%%%%%%%%%%%%%%%%%%%%%%%%%%%%
\begin{figure}%[htb!]
  \centering
  \includegraphics[width=0.8\textwidth]{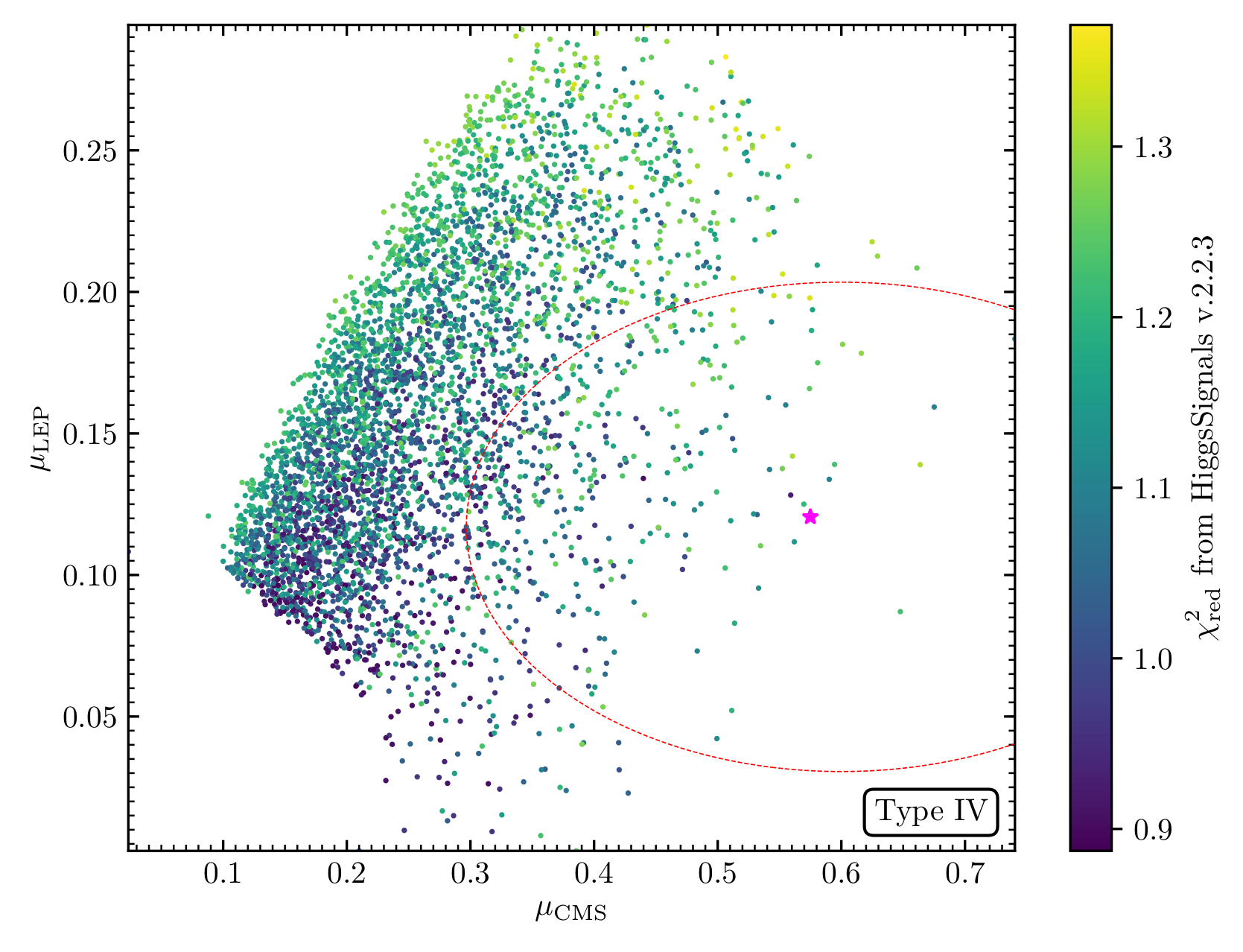}
  \vspace{-0.5cm}
  \caption{Type~IV: as in \reffi{fig:4Hpplot}, but here the colors indicate
  the $\chi_{\rm red}^2$ from HiggsSignals. The best-fit point (magenta)
  has $\chi_{\rm red}^2=1.11286$ with 101 observations considered.
  The lowest (highest) value of $\chi_{\rm red}^2$ within the
  $1 \, \sigma$ ellipse is $0.9073$ ($1.3435$).}
  \label{fig:4HSplot}
\end{figure}
%%%%%%%%%%%%%%%%%%%%%%%%%%%% F I G U R E %%%%%%%%%%%%%%%%%%%%%%%%%%%%%%

%%%%%%%%%%%%%%%%%%%%%%%%%%%% F I G U R E %%%%%%%%%%%%%%%%%%%%%%%%%%%%%%
\begin{figure}%[htb!]
  \centering
  \includegraphics[width=0.8\textwidth]{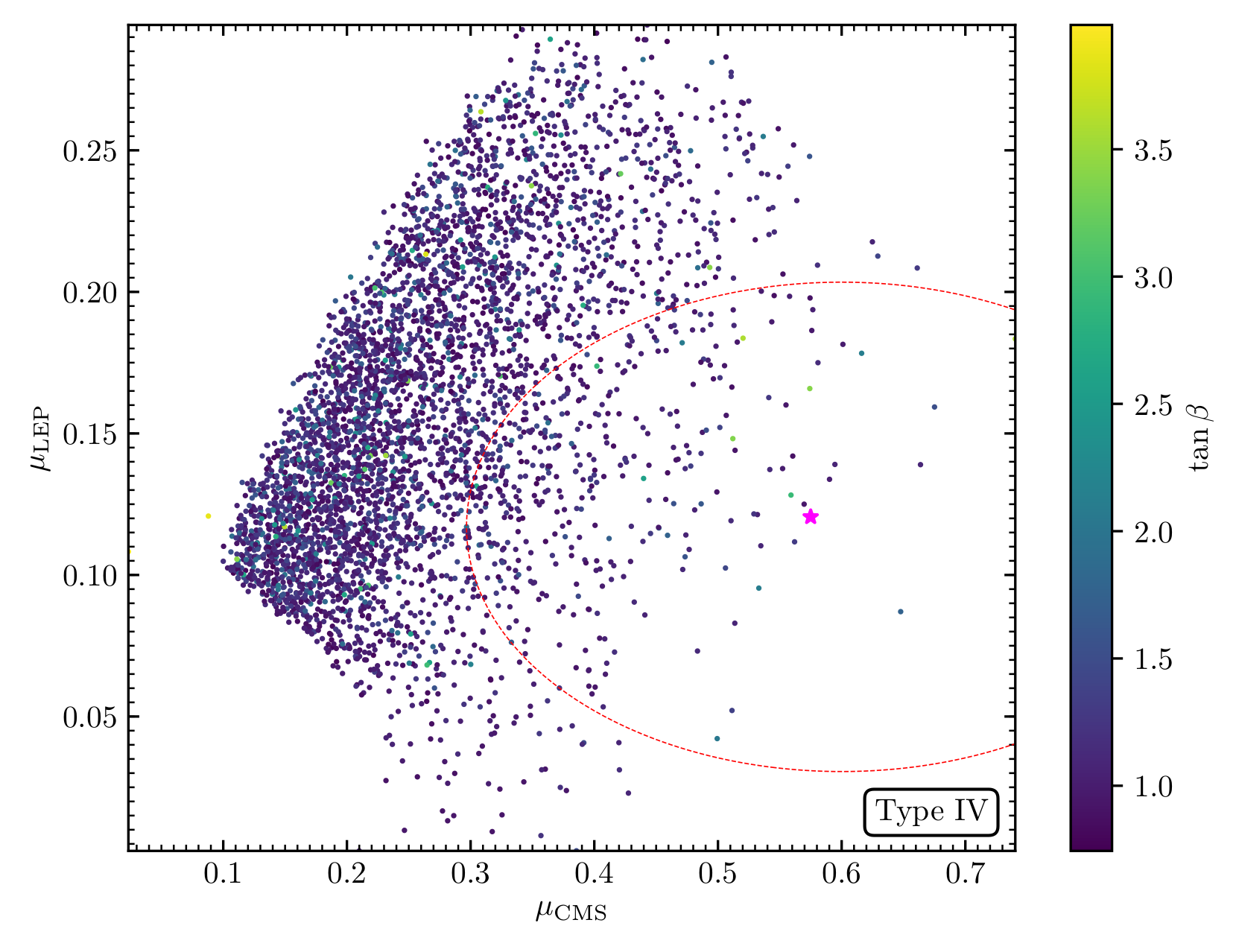}
  \vspace{-0.5cm}
  \caption{Type~IV: as in \reffi{fig:2Hpplot}, but here the colors indicate
  the value of $\tan\beta$.
  The lowest (highest) value of $\tan\beta$ within the
  $1 \, \sigma$ ellipse is $0.7935$ ($3.592$).} 
  \label{fig:4TBplot}
\end{figure}
%%%%%%%%%%%%%%%%%%%%%%%%%%%% F I G U R E %%%%%%%%%%%%%%%%%%%%%%%%%%%%%%
\begin{figure}%[htb!]
\vspace{2em}
  \centering
  \begin{subfigure}[b]{0.48\linewidth}
    \centering\includegraphics[width=\textwidth]{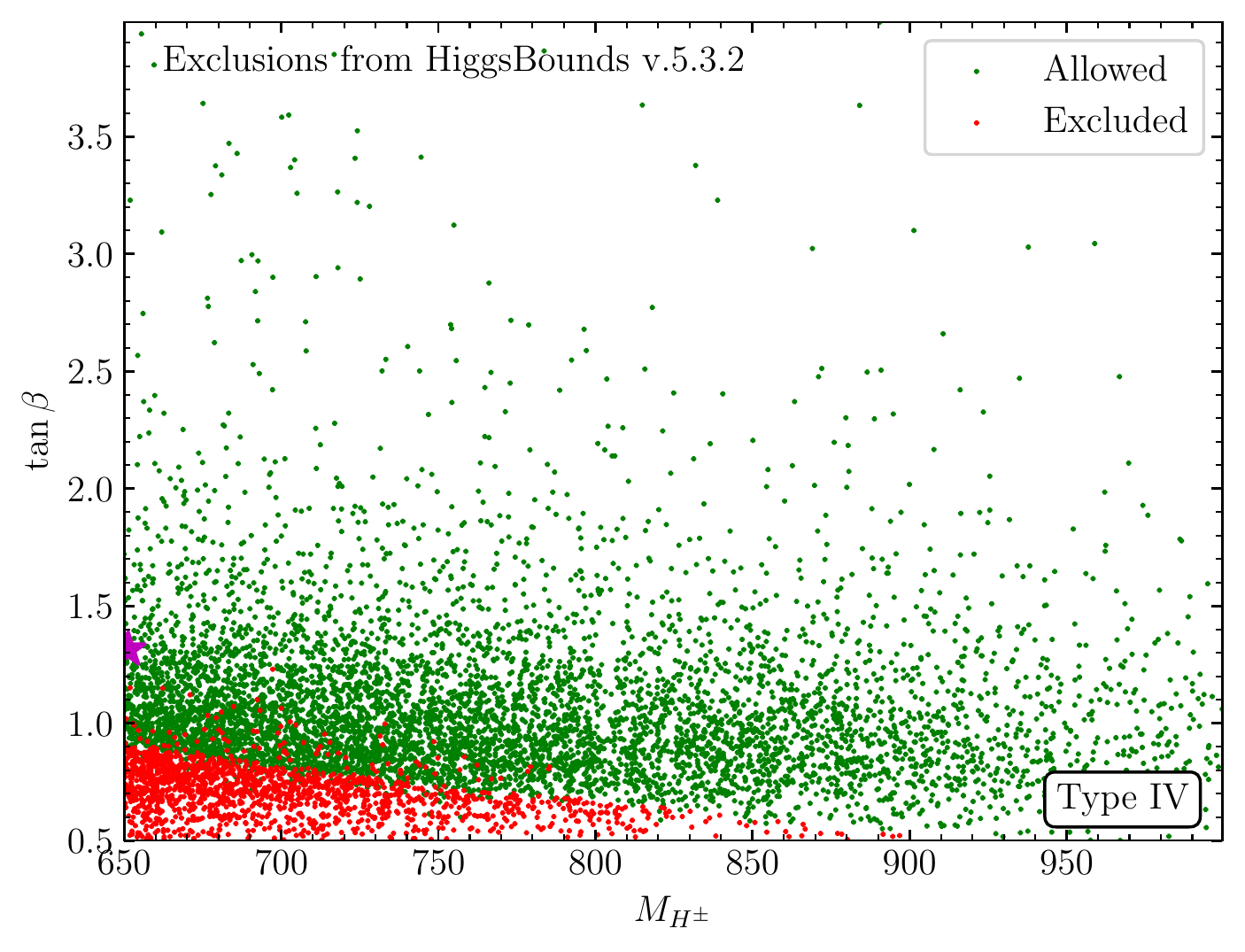}
    \caption{Direct searches at colliders}
    \label{fig:4HBplot}
  \end{subfigure}
  ~
  \begin{subfigure}[b]{0.48\linewidth}
    \centering\includegraphics[width=\textwidth]{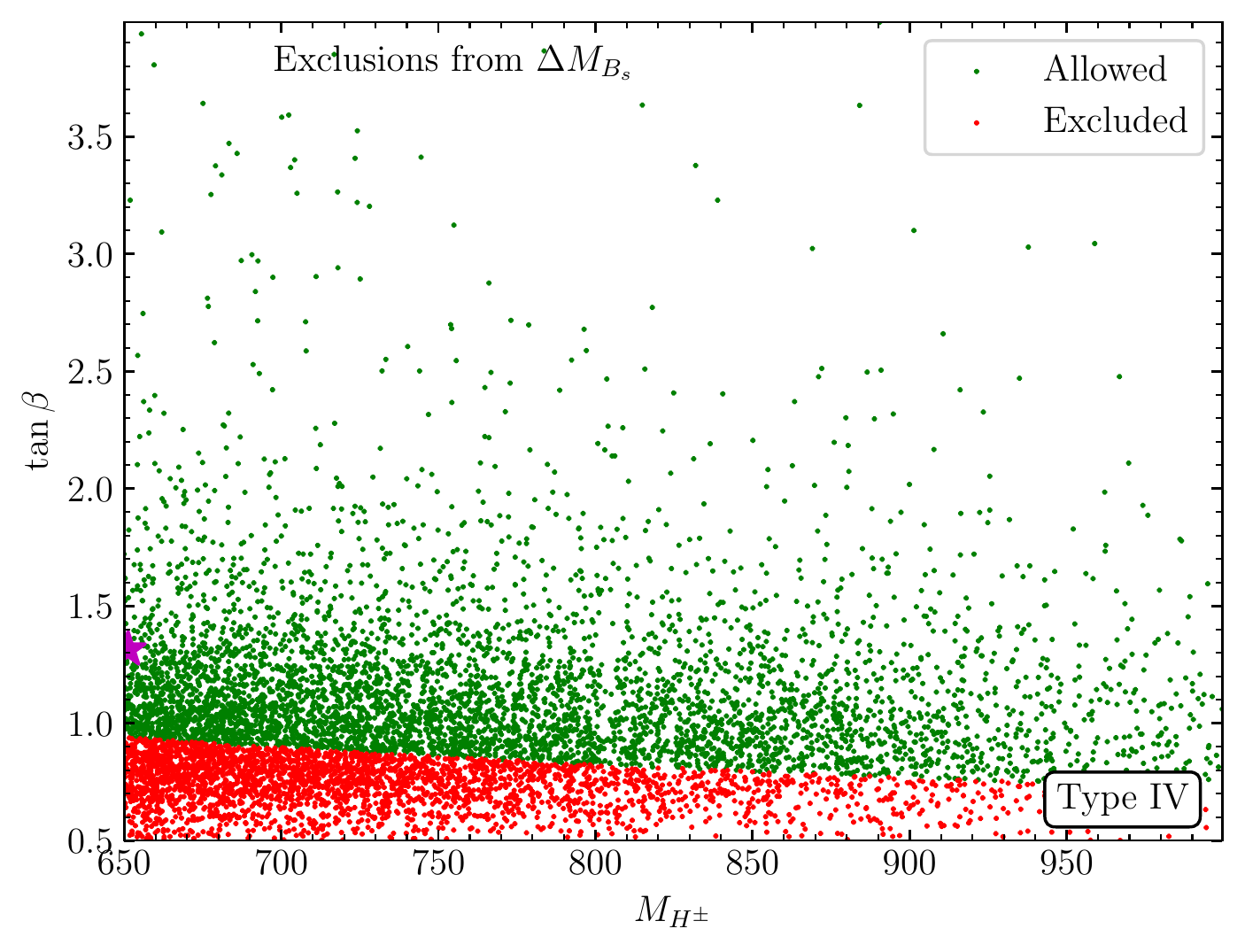}
    \caption{Flavor physics}
    \label{fig:4FLplot}
  \end{subfigure}
  \caption{Allowed (\textit{green}) and excluded 
  (\textit{red}) points 
  considering direct searches (\textit{left})
  and flavor physics (\textit{right}) in the $\MHp$-$\tb$ plane.
  The magenta star is the best-fit point.}
\end{figure}
%%%%%%%%%%%%%%%%%%%%%%%%%%%% F I G U R E %%%%%%%%%%%%%%%%%%%%%%%%%%%%%%

We show the results of our scan in the flipped scenario in
\reffis{fig:4Hpplot}-\ref{fig:4TBplot}.
Again, the color code quantifies
the charged Higgs-boson mass in \reffi{fig:4Hpplot},
the reduced $\chi^2$ from \texttt{HiggsSignals} in \reffi{fig:4HSplot},
and the value of $\tan\beta$ in \reffi{fig:4TBplot}.
As was the case in the type~II scenario, a large number of points
fit both the LEP and the CMS excesses simultaneously while being in
agreement with the measurements of the SM-like Higgs boson properties.
We again observe that the points that fit both excesses prefer low
values of $\MHp$ for the same reasons as in the type~II scenario
 (see \refse{sec:type2}).
Various points inside the $1 \, \sigma$ ellipse have additionally
a $\chi^2_{\rm red}$ from \texttt{HiggsSignals} below one, indicating
the signal strength predictions for the SM-like Higgs boson on average
are within the $1 \, \sigma$-uncertainties of each measurement.
Similar to the type~II analysis a clear preference of small $\tan\beta$-values
is visible, also for the points outside the $1 \, \sigma$ ellipse.

The exclusion boundaries from direct searches and from flavor physics
are practically the same as the ones we found in the type~II scenario.
We show in \reffi{fig:4HBplot} the allowed and excluded points of our scan
considering the collider searches in the $\tb$-$\MHp$ plane.
The most sensitive direct search is, as in type~II, the production of $H^\pm$
in association with a $tb$-pair, and subsequent decay of $H^\pm$ to
a $tb$-pair. For values of $\tb < 1$, points with a charged Higgs mass
up to $900\gev$ can be excluded.
In \reffi{fig:4FLplot} we show the excluded and allowed points regarding
constraints derived from the prediction to the meson mass difference
$\Delta M_{B_s}$. This limit is unchanged with respect to the one from
the type~II scenario, because of the similar quark Yukawa sectors in the
two cases. $\Delta M_{B_s}$ constraint is the dominant one
regarding flavor observables for the range of $\MHp$ and $\tb$ scanned here,
assuming that the exclusions from $\br(B_s \to \mu^+ \mu^-)$
constraints in the 2HDM
do not change by more than $20\%$ due to the presence of the additional
real singlet in the N2HDM~\cite{Arbey:2017gmh}.

%%%%%%%%%%%%%%%%%%%%%%%%%% T A B L E %%%%%%%%%%%%%%%%%%%%%%%%%%%%%%%%%%%%%%%%%%
\begin{table}%[htb!]
\centering
\renewcommand{\arraystretch}{1.2}
\begin{tabular}{c c c c c c c}
 $m_{h_1}$ & $m_{h_2}$ & $m_{h_3}$ & $m_A$ & $\MHp$ & & \\
 \hline
 $97.8128$ & $125.09$ & $485.998$ & $651.502$ & $651.26$ & & \\
 \hline
 \hline
 $\tb$ & $\alpha_1$ & $\alpha_2$ & $\alpha_3$ & $m_{12}^2$ & $v_S$ & \\
 \hline
 $1.3147$ & $1.27039$ & $-1.02829$ & $-1.32496$ &
   $41034.1$ & $647.886$ & \\
  \hline
  \hline
 $\br^{bb}_{h_1}$ & $\br^{gg}_{h_1}$ & $\br^{cc}_{h_1}$ &
   $\br^{\tau\tau}_{h_1}$ & $\br^{\gamma\gamma}_{h_1}$ &
     $\br^{WW}_{h_1}$ 
      & $\br^{ZZ}_{h_1}$ \\
 \hline
 $0.4074$ & $0.2071$ & $0.1189$ & $0.2483$ & $2.139\cdot 10^{-3}$ 
   & $0.0135$ 
    & $1.579\cdot 10^{-3}$ \\
 \hline
 \hline
 $\br^{bb}_{h_2}$ & $\br^{gg}_{h_2}$ & $\br^{cc}_{h_2}$ &
   $\br^{\tau\tau}_{h_2}$ & $\br^{\ga\ga}_{h_2}$ &
     $\br^{WW}_{h_2}$ & $\br^{ZZ}_{h_2}$ \\
 \hline
 $0.5363$ & $0.0939$ & $0.0345$ & $0.0758$ &
   $2.247\cdot 10^{-3}$ &  $0.2267$ & $0.0284$ \\
 \hline
 \hline
 $\br^{tt}_{h_3}$ & $\br^{gg}_{h_3}$ & $\br^{h_1 h_1}_{h_3}$ &
   $\br^{h_1 h_2}_{h_3}$ & $\br^{h_2 h_2}_{h_3}$ &
     $\br^{WW}_{h_3}$ & $\br^{ZZ}_{h_3}$ \\
 \hline
 $0.8078$ & $2.707\cdot 10^{-3}$ & $0.0124$ & $2.111\cdot 10^{-3}$ &
   $0.0119$ & $0.1085$ & $0.0517$ \\
 \hline
 \hline
 $\br^{tt}_A$ & $\br^{gg}_A$ & $\br^{Z h_1}_A$ & $\br^{Z h_2}_A$ &
   $\br^{Z h_3}_A$ & $\br^{bb}_A$ & \\
 \hline
 $0.7090$ & $1.940\cdot 10^{-3}$ & $0.1007$ & $9.652\cdot 10^{-3}$ &
   $0.1780$ & $6.49\cdot 10^{-4}$ & \\
 \hline
 \hline
 $\br^{tb}_{\Hpm}$ & $\br^{W h_3}_{\Hpm}$ & $\br^{W h_2}_{\Hpm}$ &
   $\br^{W h_1}_{\Hpm}$ & & \\
 \hline
 $0.6820$ & $0.2046$ & $9.820\cdot 10^{-3}$ & $0.1024$ & & 
\end{tabular}
\caption{Parameters of the best-fit point and branching ratios of the
scalars in the type~IV scenario.
Dimensionful parameters are given
in GeV and the angles are given in radian.}
\label{tab:4best}
\renewcommand{\arraystretch}{1.0}
\end{table}
%%%%%%%%%%%%%%%%%%%%%%%%%% T A B L E %%%%%%%%%%%%%%%%%%%%%%%%%%%%%%%%%%%%%%%%%%

The details of our best-fit point of the
scan in the N2HDM type~IV, indicated with the
magenta star in \reffis{fig:4Hpplot}-\ref{fig:4FLplot}, 
are listed in \refta{tab:4best}. The value of the charged
Higgs boson mass is just on the lower end of the scanned
range. Comparing to the best-fit point of our scan in the
N2HDM type~II, shown in \refta{tab:2best},
we observe that the values for $\tb$
and the mixing angles in the $\cp$-even scalar sector $\alpha_i$
are very similar. This is due to the fact that the effective
coefficients of the couplings of the scalars to quarks are the same.
Also the decays of the heavier Higgs bosons are similar to the
type~II best-fit point.
The striking difference between the best-fit points in both types is
that, even though the suppression of the branching ratio of $h_1$ to
$b$-quarks is larger in type~IV, the branching ratio to photons
remains smaller. As already discussed in \refse{sec:excesses},
in the parameter region, in which the excesses can be accommodated,
there is an enhancement of the decay width to $\tau$-leptons:
the value for $\br^{\tau\tau}_{h_1}$ in \refta{tab:4best} is
roughly five times larger than the one in \refta{tab:2best}.

%%%%%%%%%%%%%%%%%%%%%%%%%%%% F I G U R E %%%%%%%%%%%%%%%%%%%%%%%%%%%%%%
\begin{figure}%[htb!]
  \centering
  \begin{subfigure}[c]{0.48\linewidth}
    \centering\includegraphics[width=\textwidth]{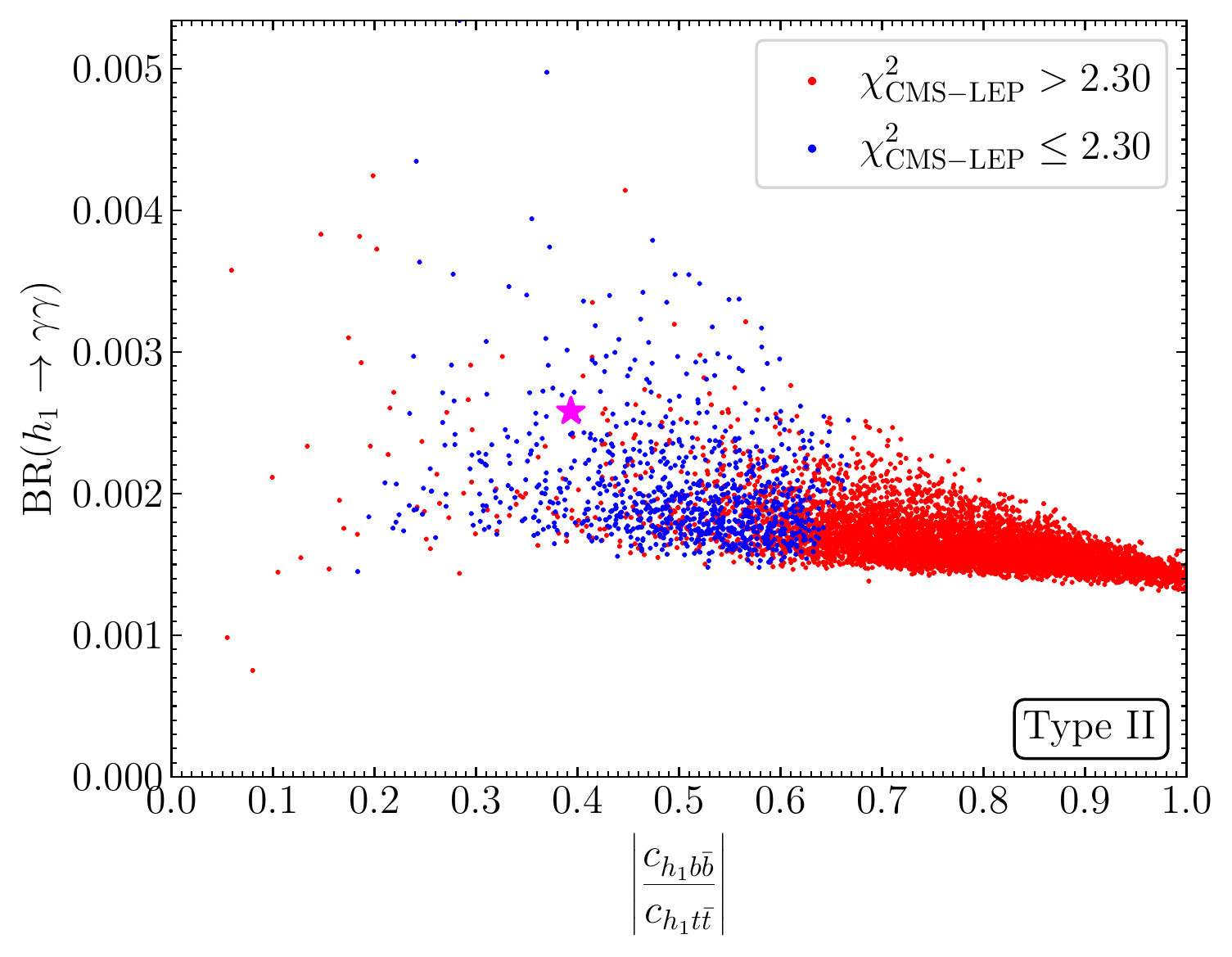}
  \end{subfigure}
  ~
  \begin{subfigure}[c]{0.48\linewidth}
    \centering\includegraphics[width=\textwidth]{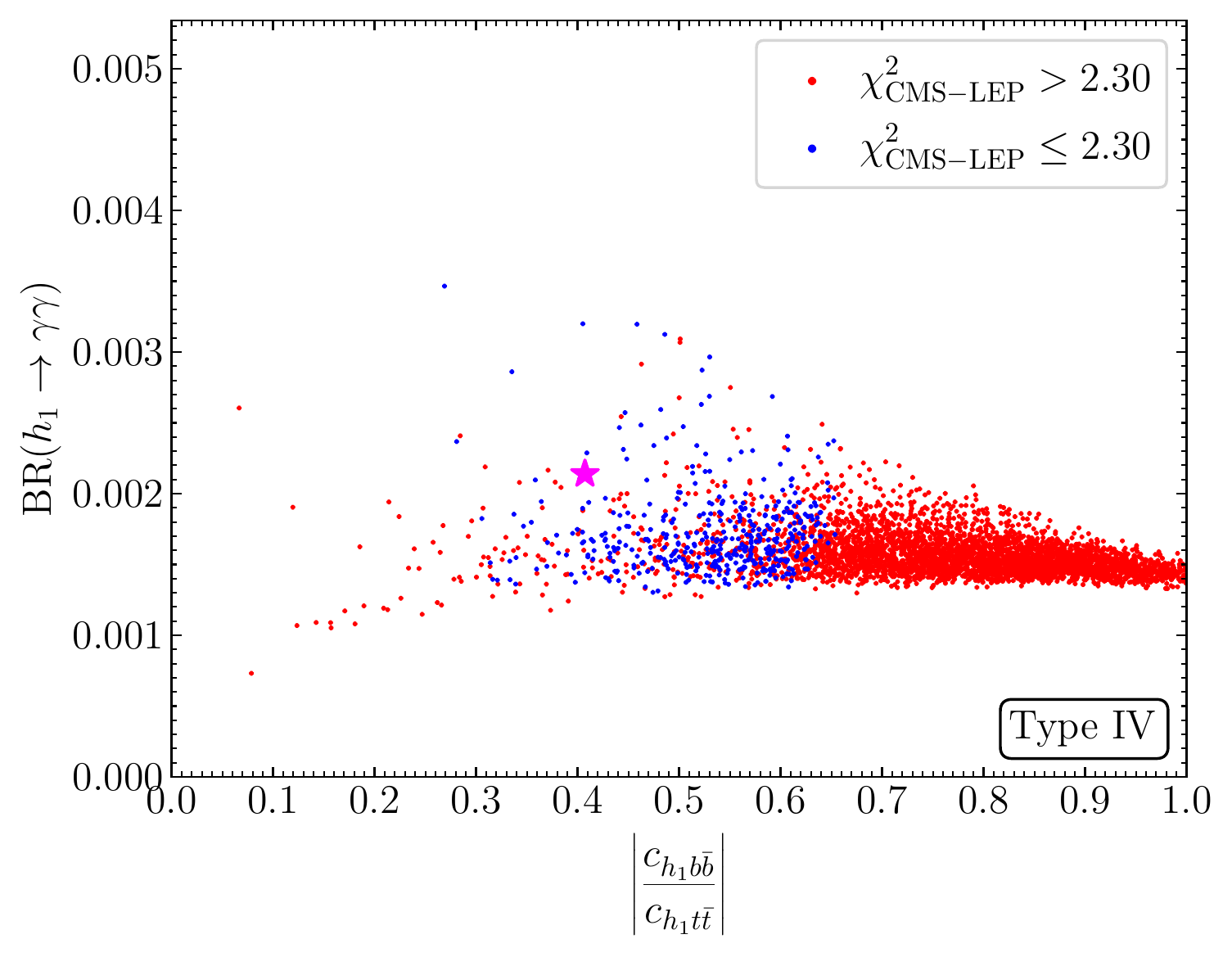}
  \end{subfigure}
  \\
  \begin{subfigure}[c]{0.48\linewidth}
    \centering\includegraphics[width=\textwidth]{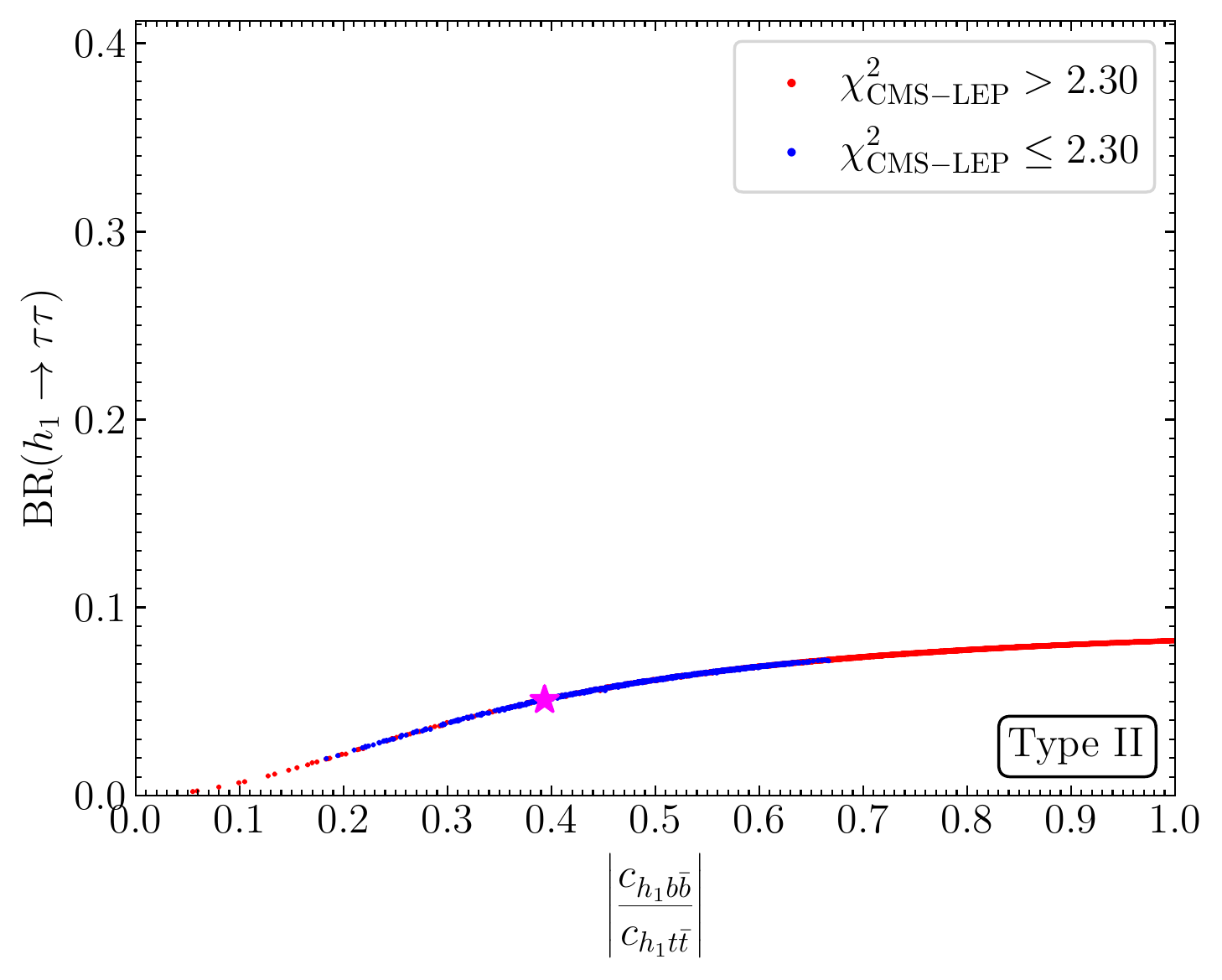}
  \end{subfigure}
  ~
  \begin{subfigure}[c]{0.48\linewidth}
    \centering\includegraphics[width=\textwidth]{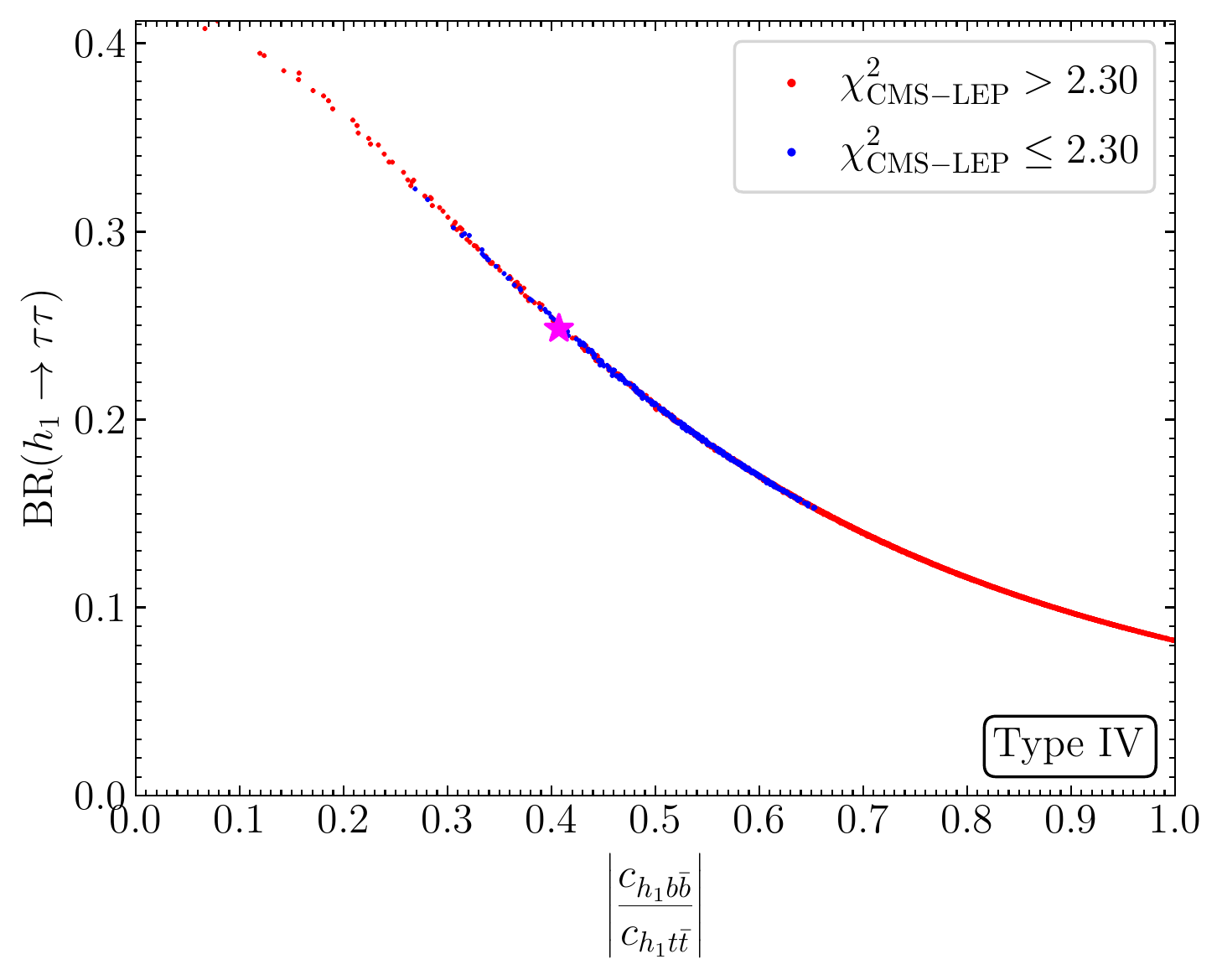}
  \end{subfigure}
  \caption{Branching fraction of $h_1$ to two photons
  (\textit{upper row})
  and to two $\tau$-leptons (\textit{lower row})
  for each parameter point respecting the experimental and
  theoretical constrains in the type~II (\textit{left})
  and the type~IV scenario
  (\textit{right}) as a function of the ratio of the coupling of $h_1$
  to bottom and top quarks normalized to the SM prediction.
  The blue points have $\chi_{\mathrm{CMS-LEP}}^2 \leq 2.30$, while the
  red points have $\chi_{\mathrm{CMS-LEP}}^2 > 2.30$.}
  \label{fig:24ctbplot}
\end{figure}
%%%%%%%%%%%%%%%%%%%%%%%%%%%% F I G U R E %%%%%%%%%%%%%%%%%%%%%%%%%%%%%%

This circumstance is not a particular feature of the best-fit point,
but a general difference between type~II and type~IV.
To illustrate this, we show in \reffi{fig:24ctbplot} the branching
ratio of $h_1$ to photons (top) and to $\tau$-leptons (bottom)
for type~II (left) and type~IV (right)
as a function of the absolute value of the ratio of the effective coupling
coefficients $c_{h_1 b \bar b}$ and $c_{h_1 t \bar t}$.
The blue and red points are the ones lying inside and outside
the $1 \, \sigma$ ellipse regarding $\chi_{\mathrm{CMS-LEP}}^2$, respectively.
When $|c_{h_1 b \bar b}/c_{h_1 t \bar t}|$ is small, the branching
ratios to photons receives 
an enhancement and it is possible to fit the CMS excess.
However, in type~II the enhancement is larger
than in type~IV, because the branching
ratio to $\tau$-leptons scales with the same factor as $c_{h_1 b \bar b}$
in type~II, but proportional to $c_{h_1 t \bar t}$ in type~IV.

\begin{figure}
  \centering
  \begin{subfigure}[b]{0.48\linewidth}
    \centering\includegraphics[width=\textwidth]{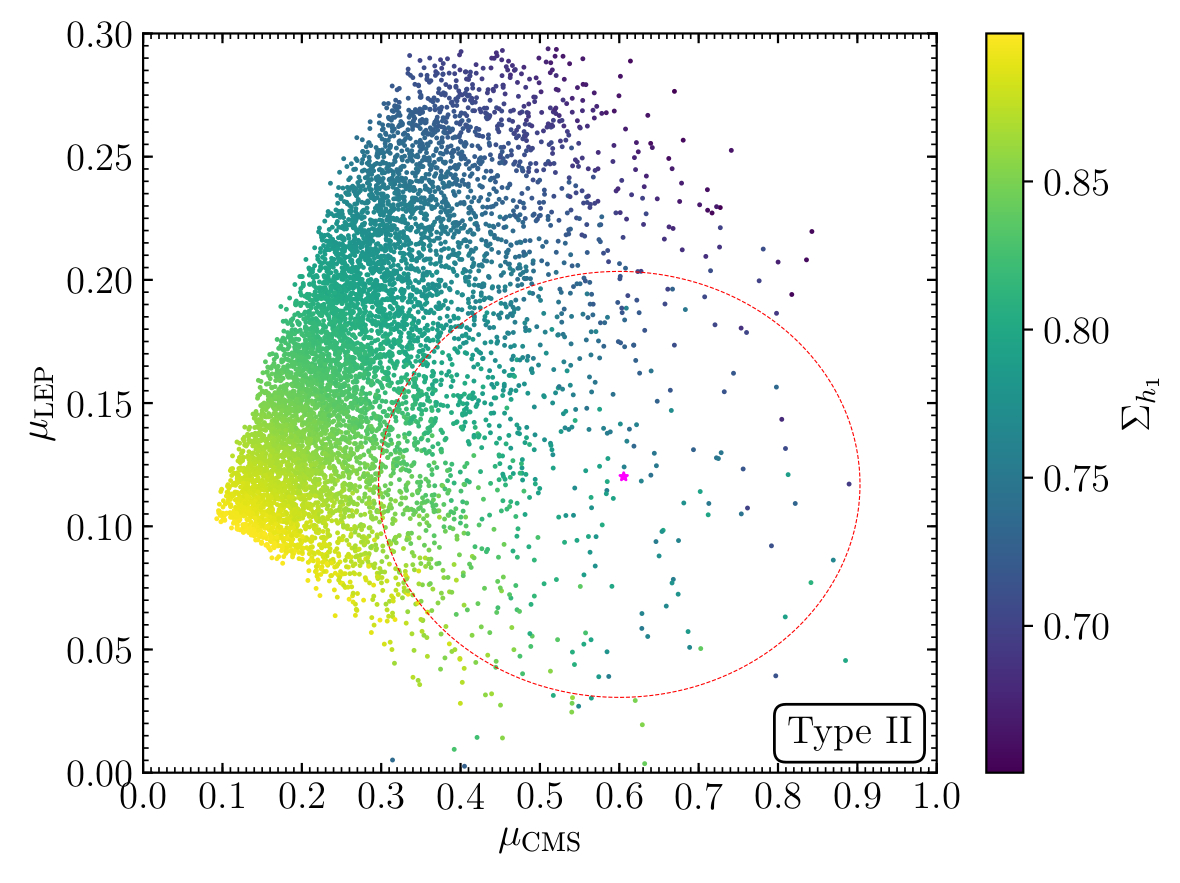}
    \caption{Type~II}
  \end{subfigure}
  ~
  \begin{subfigure}[b]{0.48\linewidth}
    \centering\includegraphics[width=\textwidth]{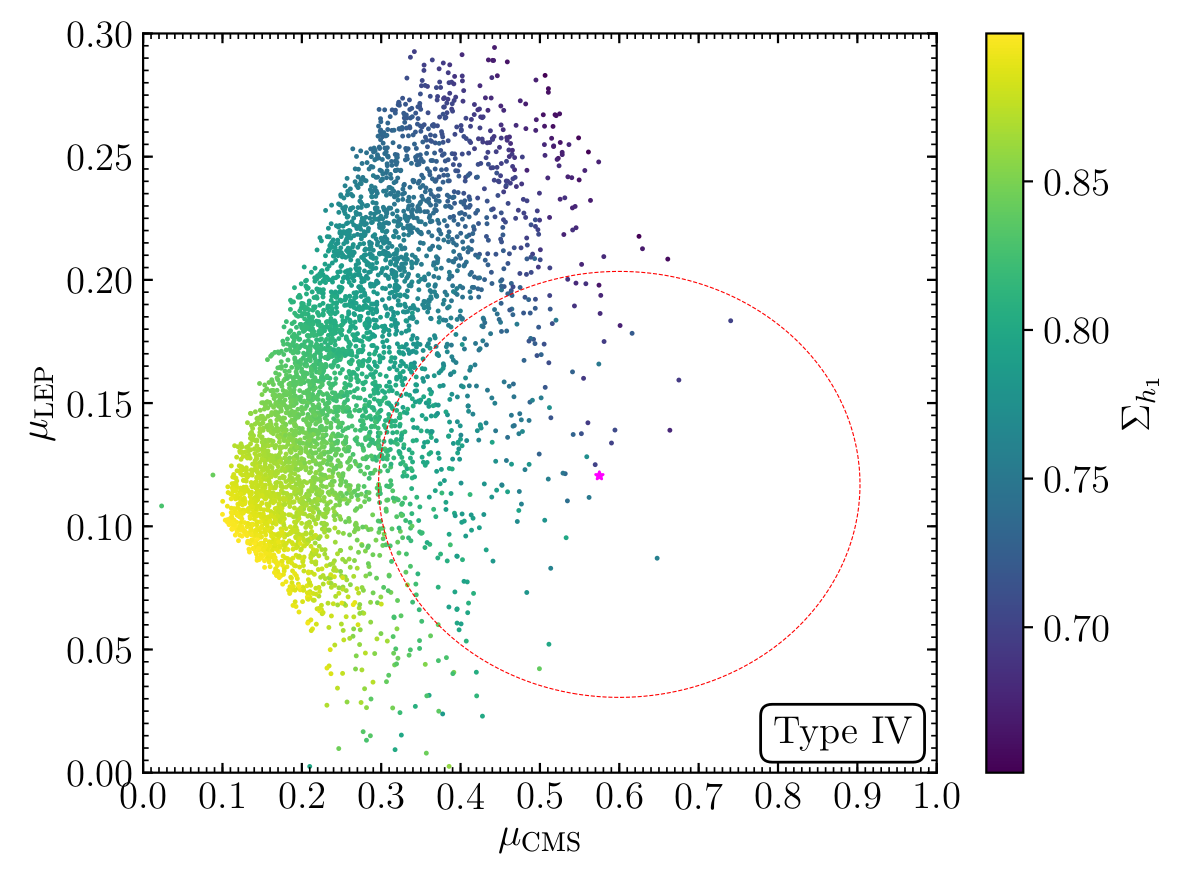}
    \caption{Type~IV (flipped)}
  \end{subfigure}
  \caption{Shown are the signal strengths $\mu_{\rm CMS}$ and $\mu_{\rm LEP}$
  for each parameter point respecting the experimental and
  theoretical constrains in the type~II and the type~IV scenario.
  The $1 \, \sigma$-region of both excesses
  is shown by the red ellipse. The colors show the singlet
  component of $h_1$. The magenta star is the best-fit point.}
  \label{fig:ty2ty4sH1}
\end{figure}

In \reffi{fig:ty2ty4sH1} we show the signal strengths for both excesses
in the N2HDM type~II (left) and type~IV (right), with colors
indicating the singlet component of $h_1$.
Comparing both plots, it becomes apparent that the effect mentioned
above results in a substantial suppression of $\mu_{\rm CMS}$ in the
type~IV scenario. For similar values of the singlet component
$\Sigma_{h_1}$, the type~II scenario can reach larger $\mu_{\rm CMS}$,
whereas the size of $\mu_{\rm LEP}$ is very similar in both scenarios.
Remarkably, the type~II scenario can reach values of $\mu_{\rm CMS} \sim 1$,
meaning that the signal strength prediction for $\mu_{\rm CMS}$ is as big
as the one of a hypothetical SM-like Higgs boson at the same mass,
even though it is dominantly singlet-like.
In the type~IV scenario, on the other hand, there is no point above the
upper $1 \, \sigma$-limit of $\mu_{\rm CMS}=0.8$.
As one can anticipate form these plots, points with
$\Sigma_{h_1} \ge 0.9$ are not expected in the $1\,\sig$ ellipse. We
have verified this by dedicated scans, i.e.\ it is confirmed that
$\Sigma_{h_1} \le 0.9$ does not have a relevant impact on the overall
results of our analysis.

%%%%%%%%%%%%%%%%%%%%%%%%%%%% F I G U R E %%%%%%%%%%%%%%%%%%%%%%%%%%%%%%
%%%%%%%%%%%%%%%%%%%%%%%%%%%% F I G U R E %%%%%%%%%%%%%%%%%%%%%%%%%%%%%%

%%%%%%%%%%%%%%%%%%%%%%%%%%%%%%%%%%%%%%%%%%%%%%%%%%%%%%%%%%%%%%%%%%%%%%%%%%

\subsection{Future searches}
\label{sec:future}

A light singlet-like light scalar, as is present in the N2HDM, is 
very challenging to directly search for at the LHC, 
because of its suppressed couplings to all SM particles. 
That is why it might have escaped discovery so far
except for some alluring hints two of which 
we have focussed on in this work.
Indirect probes for such a particle are possible with
precision measurements of the couplings of the $125\gev$
Higgs state.
We will discuss both possibilities as well as searches for heavy
Higgs bosons in the following.

%%%%%%%%%%%%%%%%%%%%%%%%%%%%%%%%%%%%%%%%%%%%%%%%%%%%%%%%%%%%%%%%%%%%%%%%%%%%%%%

\subsubsection{Indirect searches}
\label{sec:propsindirect}

Currently, uncertainties on the measurement of the coupling
strengths of the SM-like Higgs boson at the LHC are still large,
i.e., at the $1 \, \sigma$-level they are of the same order as the
modifications of the couplings present
in our analysis in the
N2HDM~\cite{Khachatryan:2016vau,ATLAS-CONF-2018-031,Sirunyan:2018koj}.
In the future, once the complete $300\, \ifb$ collected at the LHC are analyzed,
the constraints on the couplings of the SM-like
Higgs boson will benefit from the reduction of statistical uncertainties.
Even tighter constraints are expected from the LHC after the
high-luminosity upgrade (HL-LHC), when the planned amount
of $3000\, \ifb$ integrated luminosity will have been collected~\cite{Dawson:2013bba}.
Finally, a future linear $e^+ e^-$ collider like the ILC could
improve the precision measurements of the Higgs boson couplings
even further~\cite{Dawson:2013bba,Drechsel:2018mgd}.\footnote{Similar
results can be obtained for CLIC, FCC-ee and CEPC. We will focus on the ILC 
prospects here using the results of \citere{Drechsel:2018mgd}.} Firstly, a lepton collider has the
advantage of massively reduced QCD background compared to a
hadron collider like the LHC.
Secondly,
the cross section of the 
Higgs boson can be measured independently,
and the total width (and therefore also the coupling
modifiers) can be reconstructed without model assumptions.

Several studies have been performed to estimate the future constraints
on the coupling modifiers of the SM-like Higgs boson at the
LHC~\cite{Dawson:2013bba,CMS:2013xfa,Tricomi:2015nrd,ATL-PHYS-PUB-2014-016,Slawinska:2016zeh} and the
ILC~\cite{Dawson:2013bba,Asner:2013psa,Ono:2013sea,Durig:2014lfa,Fujii:2017vwa,Bambade:2019fyw,Cepeda:2019klc},
assuming that no deviations from the SM predictions will be found.
Here, we illustrate the capability of both experiments to either rule
out or confirm the scenarios we presented in our paper.
We compare our scan points to the expected precisions of
the LHC and the ILC as they are reported
in \citeres{Bambade:2019fyw,Cepeda:2019klc},
neglecting possible correlations of the coupling modifiers.

%%%%%%%%%%%%%%%%%%%%%%%%%%%% F I G U R E %%%%%%%%%%%%%%%%%%%%%%%%%%%%%%
\begin{figure}
  \centering
  \begin{subfigure}[c]{0.6\textwidth}
    \centering\includegraphics[width=\textwidth]{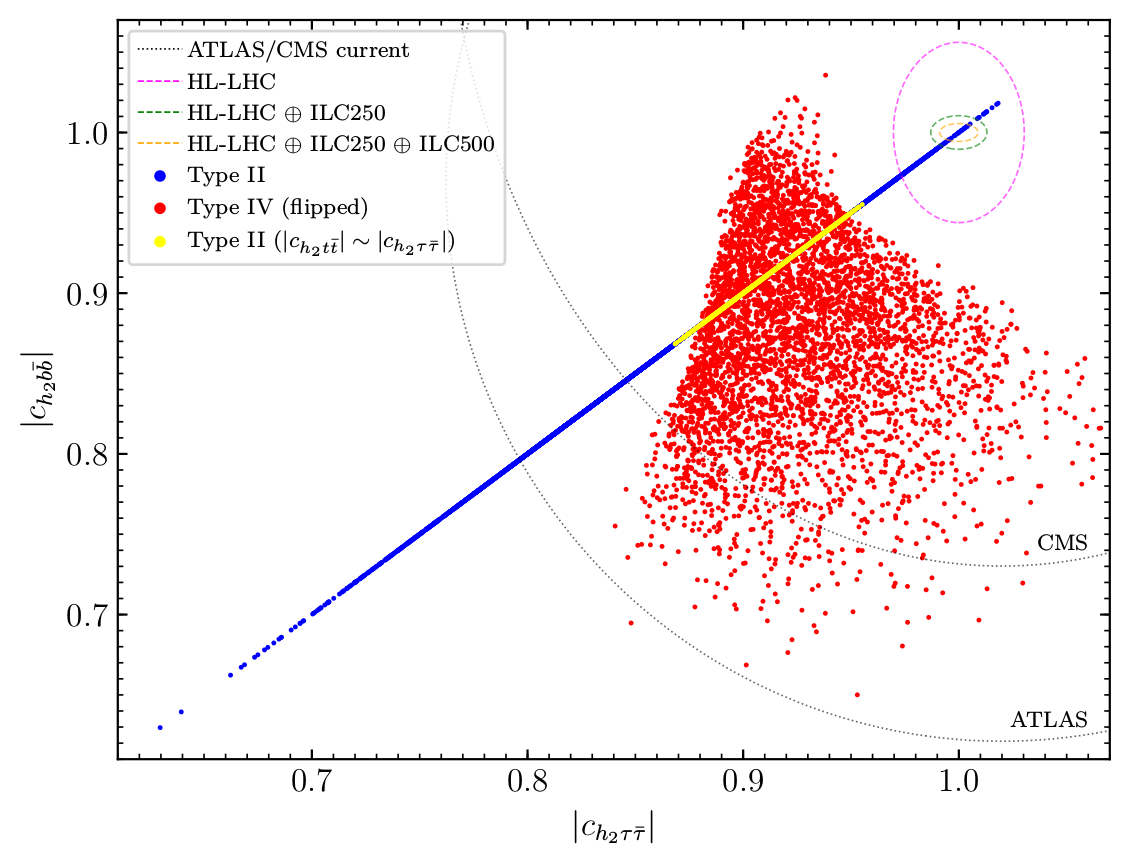}
  \end{subfigure}
  \begin{subfigure}[c]{0.6\textwidth}
     \centering\includegraphics[width=\textwidth]{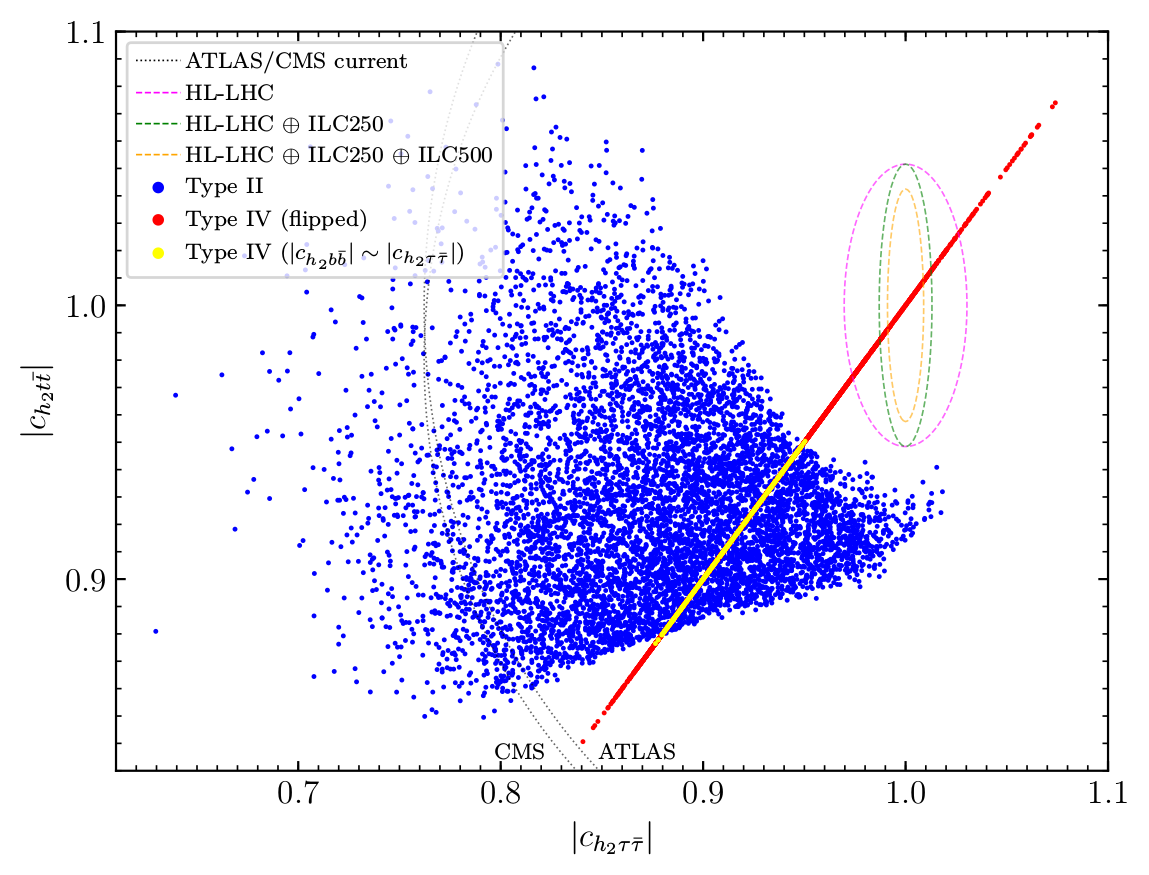}
  \end{subfigure}
  \caption{Scan points of our analysis in the
  type~II (\textit{blue}) and type~IV (\textit{red}) scenario in the
  $|c_{h_2 \tau \bar{\tau}}|$-$|c_{h_2 b \bar{b}}|$ plane (\textit{top}) and
  the $|c_{h_2 \tau \bar{\tau}}|$-$|c_{h_2 t \bar{t}}|$ plane (\textit{bottom}).
  In the upper plot we highlight in yellow the points of the type~II
  scenario that overlap with points from the type~IV scenario
  in the lower plot, i.e., points with
  $|c_{h_2 t \bar{t}}| \sim |c_{h_2 b \bar{b}}| \sim |c_{h_2 \tau \bar{\tau}}|$.
  In the same way in the lower plot we highlight
  in yellow the points of the type~IV
  scenario that overlap with points from the type~II scenario
  in the upper plot.
  The dashed ellipses are the projected uncertainties
  at the HL-LHC~\cite{Cepeda:2019klc} (\textit{magenta}) and the
  ILC~\cite{Bambade:2019fyw} (\textit{green} and \textit{orange})
  of the measurements of
  the coupling modifiers at the $68\%$ confidence level, assuming
  that no deviation from the SM prediction will be found
  (more details in the text).
  We also show with the dottet black lines the $1 \, \sigma$ ellipses
  of the current measurements from
  CMS~\cite{Sirunyan:2018koj} and
  ATLAS~\cite{ATLAS-CONF-2018-031}.}
  \label{fig:cplprosp}
\end{figure}
%%%%%%%%%%%%%%%%%%%%%%%%%%%% F I G U R E %%%%%%%%%%%%%%%%%%%%%%%%%%%%%%

In \reffi{fig:cplprosp} we plot the effective
coupling coefficient of the SM-like Higgs boson $h_2$ to $\tau$-leptons
on the horizontal axis against the coupling coefficient to
$b$-quarks (top) and to $t$-quarks (bottom) for both types.
These points passed all the experimental and theoretical constraints, including
the verification of SM-like Higgs boson properties in agreement
with LHC results using \texttt{HiggsSignals}.
In the top plot the blue points lie on a diagonal line, because in
type~II the coupling to leptons and to down-type quarks scale identically,
while in the bottom plot the red points representing
the type~IV scenario lie on the diagonal,
because there the lepton-coupling scales in the same
way as the coupling to up-type quarks.
The current measurements
on the coupling modifiers by ATLAS~\cite{ATLAS-CONF-2018-031}
and CMS~\cite{Sirunyan:2018koj} are shown
as black ellipses, although
the corresponding uncertainties are still very large.

The magenta ellipse in each plot shows the expected precision of the measurement
of the coupling coefficients at the $1 \, \sigma$-level at the
HL-LHC from \citere{Cepeda:2019klc}.
The current uncertainties and the HL-LHC analysis are based on the
coupling modifier, or $\kappa$-framework, in which the tree-level
couplings of the SM-like Higgs boson to
vector bosons, the top quark, the bottom quark, the $\tau$ and the $\mu$ lepton,
and the three loop-induced
couplings to $\ga\ga$, $gg$ and $Z\gamma$ receive a factor $\kappa_i$ quantifying
potential modifications from the SM predictions. These modifiers are then
constrained using a global fit to projected HL-LHC data assuming no
deviation from the SM prediction will be found. The uncertainties found
for the $\kappa_i$ can directly be applied to the future precision
of the coupling modifiers $c_{h_i \dots}$ we use in our paper.
We use the uncertainties given under the assumptions that
no decay of the SM-like Higgs boson to BSM particles is present,
and that current systematic uncertainties will be reduced in addition
to the reduction of statistical uncertainties due to the increased statistics.

The green and the orange ellipses show the corresponding expected
uncertainties when the HL-LHC results are combined with projected
data from the ILC after the $250\gev$ phase and
the $500\gev$ phase, respectively, taken from \citere{Bambade:2019fyw}.
Their analysis is based on a pure effective field theory calculation,
supplemented by further assumptions to facilitate the combination with
the HL-LHC projections in the $\kappa$-framework. In particular, in
the effective field theory approach the vector boson couplings can
be modified beyond a simple rescaling. This possibility was excluded
by recasting the fit setting two parameters related to the couplings
to the $Z$-boson and the $W$-boson to zero
(for details we refer to \citere{Bambade:2019fyw}).

Remarkably, while current constraints on the SM-like Higgs-boson
properties allow for large deviations of the couplings of up
to $40\%$, the parameter space of our scans will be significantly
reduced by the expected constraints from the HL-LHC
and the ILC. 
For instance, the uncertainty of the coupling to $b$-quarks
will shrink below $4\%$ at the HL-LHC\footnote{Here one has to
keep in mind the theory input required in the (HL-)LHC analysis.}
and below $1\%$ at the ILC.
For the coupling to $\tau$-leptons the uncertainty is expected to be
at $2\%$ at the HL-LHC. Again, the ILC could reduce this uncertainty further
to below $1\%$. For the coupling to $t$-quarks, on the other
hand, the ILC cannot improve substantially the expected uncertainty of
the HL-LHC (but permit a model-independent analysis).
Still, the HL-LHC and the ILC are
expected to reduce the uncertainty by roughly a factor of three.
This demonstrates that our explanation of the LEP
and the CMS excesses within the N2HDM is testable indirectly using future
precision measurements of the SM-like Higgs-boson couplings.
Independent of the type of the N2HDM, we can see comparing both
plots in \reffi{fig:cplprosp}, that there is not a single benchmark
point that coincides with the SM prediction regarding the three
coupling coefficients shown. This implies that, once these
couplings are measured precisely by the HL-LHC and the ILC,
a deviation of the SM prediction has to be measured in at least one
of the couplings, if our explanation of the excesses is correct.
Accordingly, if no deviation from the SM prediction regarding these
couplings will be measured, our explanation would be ruled out entirely.
Of course, this result is not surprising, as we explicitly
demanded a lower limit on the singlet component of the SM-like Higgs boson
of $\Sigma_{h_2} \geq 10\%$ in our scans.
Consequently, the second lightest Higgs boson naturally exhibits
deviations regarding the coupling coefficients.
However, we checked explicitely by dedicated scans, as discussed above
that benchmark points with $\Sigma_{h_2} < 10\%$
cannot accommodate both excesses, because in that case the doublet
component of $h_1$ is too small. Thus,
the conclusions w.r.t.\ the coupling analysis is not affected by
demanding $\Sigma_{h_2} \ge 0.1$.

Furthermore, in case a deviation from the SM prediction will be found,
the predicted scaling behavior of the coupling coefficients in the type
II scenario (upper plot) and the type~IV scenario (lower plot), might
lead to distinct possibilities for the two models to accommodate these
possible deviations. In this case, precision measurements of the
SM-like Higgs boson couplings could be used to exclude one of the
two scenarios.
This is true for all points except the ones highlighted
in yellow in \reffi{fig:cplprosp}. The yellow points are a subset of points
of our scans that, if such deviations of the SM-like Higgs boson
couplings will be measured, could correspond to a benchmark point
of both the scan in the type~II and the type~IV scenario.
However, note that this subset of points is confined to the diagonal
lines of both plots, and thus corresponds to a very specific subset
of the overall allowed parameter space.
For the type~II scenario, in the upper plot,
the yellow points are determined by the additional
constraint that $|c_{h_2 t \bar{t}}| \sim |c_{h_2 \tau \bar{\tau}}|$,
which is exactly true in the type~IV scenario.
For the type~IV scenario, in the lower plot,
the yellow points are determined by the additional
constraint that $|c_{h_2 b \bar{b}}| \sim |c_{h_2 \tau \bar{\tau}}|$,
which is exactly true in the type~II scenario.

%%%%%%%%%%%%%%%%%%%%%%%%%%%% F I G U R E %%%%%%%%%%%%%%%%%%%%%%%%%%%%%%
\begin{figure}
  \centering
  \begin{subfigure}[c]{0.6\textwidth}
    \centering\includegraphics[width=\textwidth]{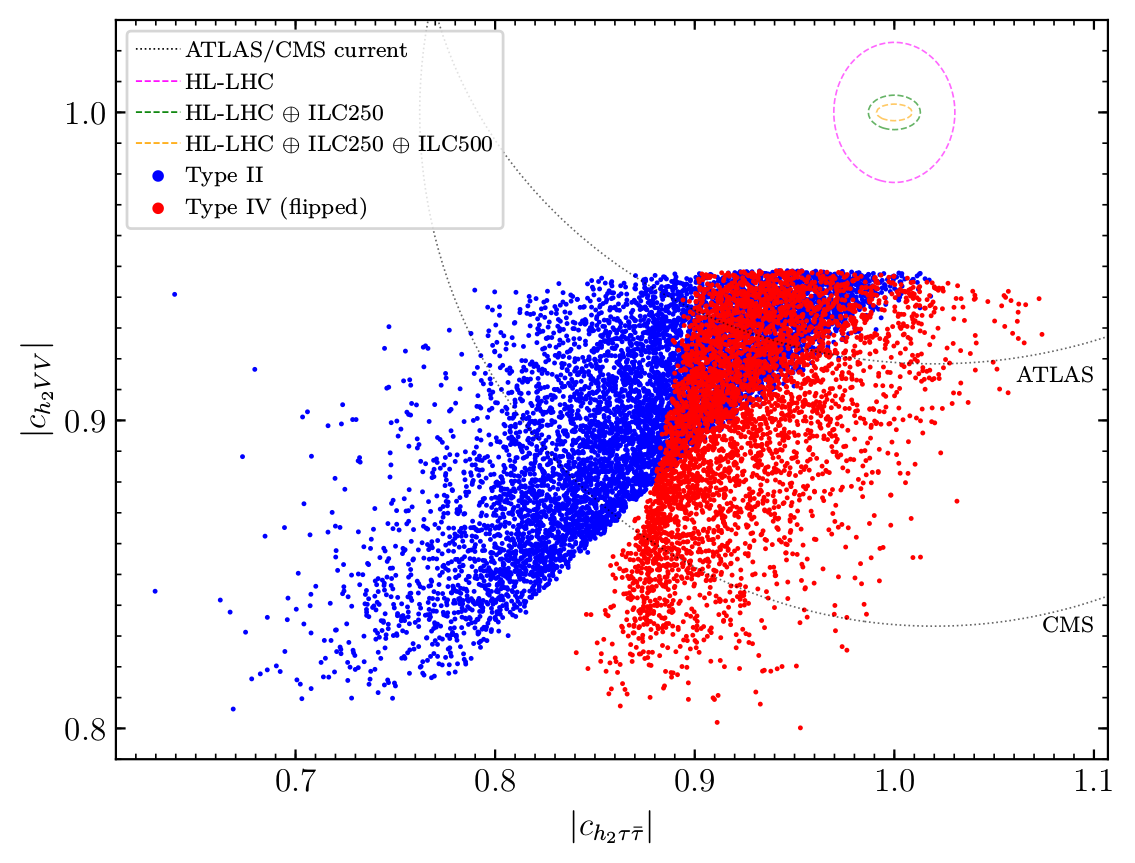}
  \end{subfigure}
  \caption{As in \reffi{fig:cplprosp} but with $|c_{h_2 V V}|$
  on the vertical axis.}
  \label{fig:cplprospvv}
\end{figure}
%%%%%%%%%%%%%%%%%%%%%%%%%%%% F I G U R E %%%%%%%%%%%%%%%%%%%%%%%%%%%%%%

For completeness we show in \reffi{fig:cplprospvv} the
absolute value of the coupling
modifier of the SM-like Higgs boson w.r.t.\ the vector boson couplings
$|c_{h_2 V V}|$ on the vertical axis. Again, the parameter points of both
types show deviations larger than the projected experimental uncertainty
at HL-LHC and ILC.
The deviations in $|c_{h_2VV}|$ are even stronger than for the
couplings to fermions. A $2\,\sig$ deviation from the SM prediction is expected
with HL-LHC accuracy. At the ILC a deviation fo more than $5\,\sig$
would be visible.
As mentioned already, a suppression of the
coupling to vector bosons is explicitly
expected by demanding $\Sigma_{h_2} \geq 10\%$. However,
since points with lower singlet component cannot accommodate
both excesses, this does not contradict the conclusion that
the explanation of both excesses can be
probed with high significance with future Higgs-boson coupling
measurements.

%%%%%%%%%%%%%%%%%%%%%%%%%%%%%%%%%%%%%%%%%%%%%%%%%%%%%%%%%%%%%%%%%%%%%%%%%%%%%%%

\subsubsection{Direct searches}
\label{sec:propsdirect}

Direct searches for the singlet-dominated scalar is particularly
challenging at the LHC due to the large background, especially since the
mass scale is close to the $Z$-boson resonance. In spite of that,
the diphoton bump which has persisted through LHC Run\,I and II
is worth exploring in additional Higgs boson searches of future runs of the LHC.
In particular the search for charged Higgs bosons appears promising in the
region of low $\tb$. In \refse{subsec:collider} we have
indicated that indeed already
with the current data the charged Higgs-boson searches
with $H^\pm \to tb$ provide an important
constraint in the favored region of parameter space. Consequently, further
searches at the (HL-)LHC will yield stronger constraints or (hopefully)
discover signs of a charged Higgs boson in the region between $600 \gev$ and
$950 \gev$. Prospects for a $5\,\sig$ discovery
in the charged Higgs-boson searches can be
found in \citere{Guchait:2018nkp}.
The prospects for the searches for the heavy neutral Higgs bosons,
decaying dominantly to $t \bar t$, may also be promising. However, we
are not aware of corresponding HL-LHC projections.

%%%%%%%%%%%%%%%%%%%%%%%%%%%% F I G U R E %%%%%%%%%%%%%%%%%%%%%%%%%%%%%%
\begin{figure}%[htb!]
  \centering
  \begin{subfigure}[c]{0.8\linewidth}
    \centering\includegraphics[width=\textwidth]{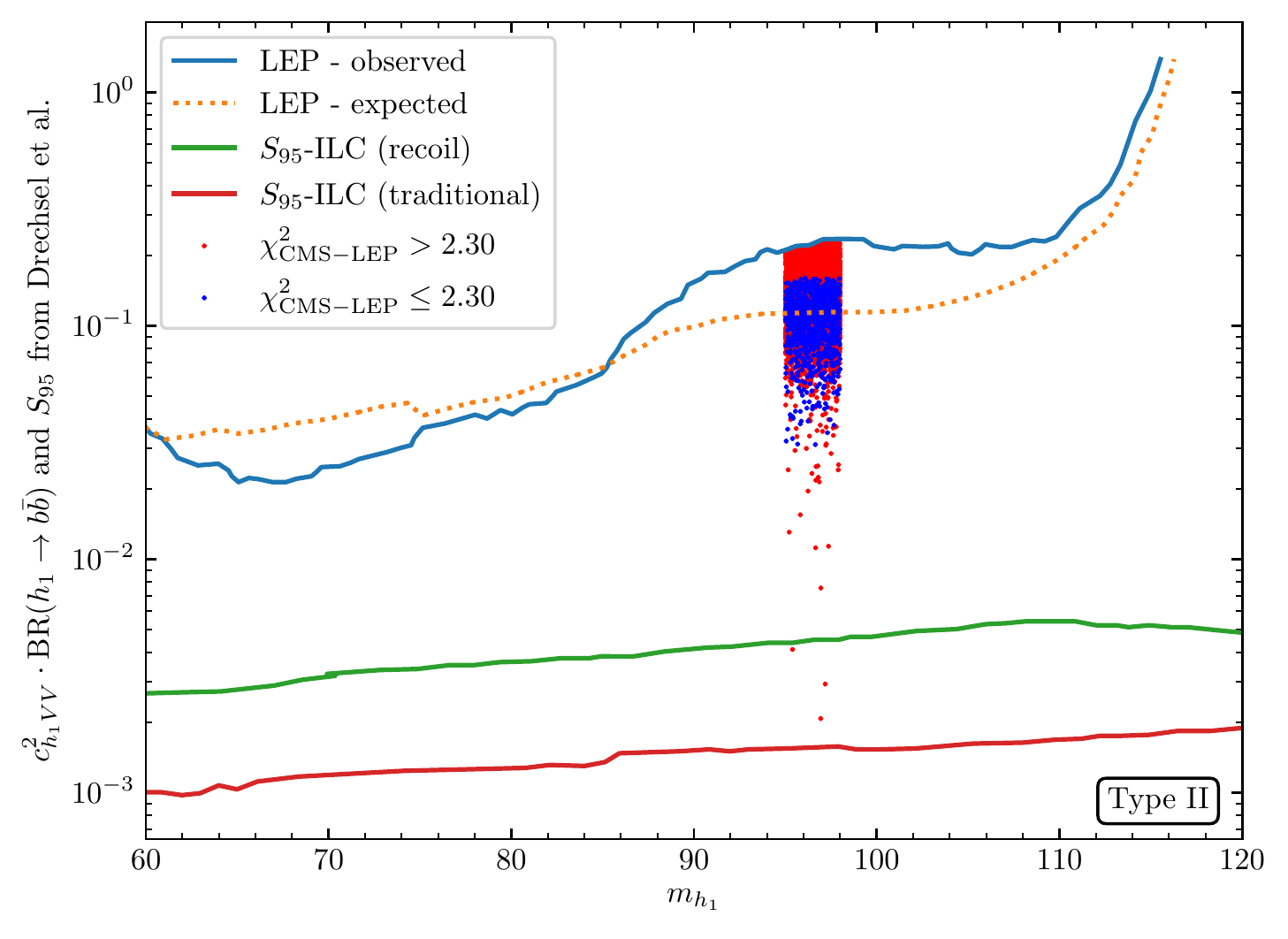}
  \end{subfigure}
  \caption{The $95\%$ CL expected (\textit{orange dashed}) and
  observed (\textit{blue}) upper bounds on the Higgsstrahlung production
  process with associated decay of the scalar to a pair of
  bottom quarks at LEP~\cite{Barate:2003sz}.
  Expected $95\%$ CL upper limits on the Higgsstrahlung production
  process normalized to the SM prediction $S_{95}$ at
  the ILC using the traditional (\textit{red}) and the
  recoil technique (\textit{green}) as described in the
  text~\cite{Drechsel:2018mgd}. We also show the points of our scan
  in the type~II scenario which lie within (\textit{blue}) and
  outside (\textit{red}) the $1 \, \sigma$ ellipse
  regarding the CMS and the LEP
  excesses.}
  \label{fig:2ilc}
\end{figure}
%%%%%%%%%%%%%%%%%%%%%%%%%%%% F I G U R E %%%%%%%%%%%%%%%%%%%%%%%%%%%%%%

%%%%%%%%%%%%%%%%%%%%%%%%%%%% F I G U R E %%%%%%%%%%%%%%%%%%%%%%%%%%%%%%
\begin{figure}%[htb!]
  \centering
  \begin{subfigure}[c]{0.8\linewidth}
    \centering\includegraphics[width=\textwidth]{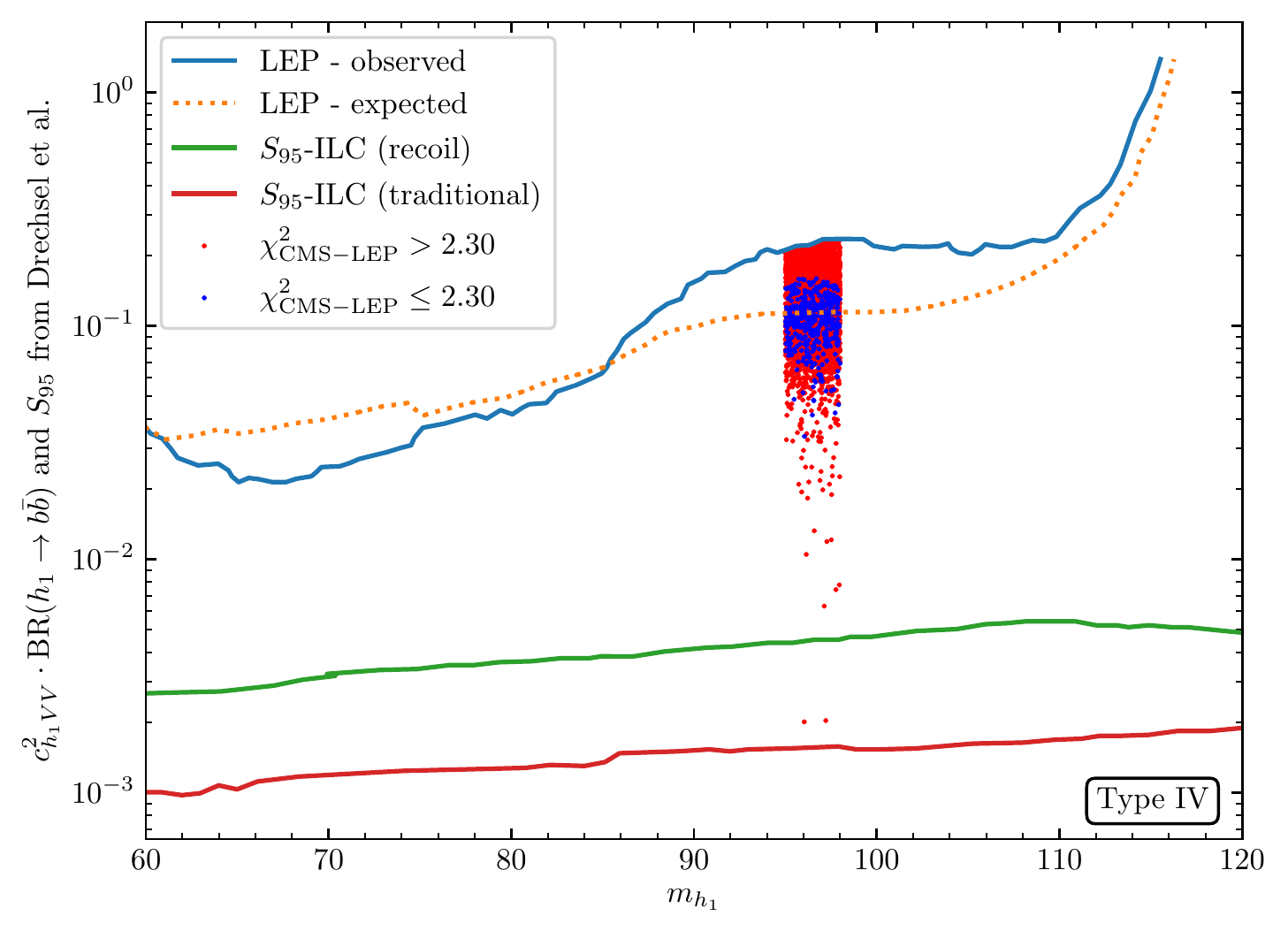}
  \end{subfigure}
  \caption{The same as in \reffi{fig:2ilc}, but with the points
  of our scan in the type~IV scenario.}
  \label{fig:4ilc}
\end{figure}
%%%%%%%%%%%%%%%%%%%%%%%%%%%% F I G U R E %%%%%%%%%%%%%%%%%%%%%%%%%%%%%%

$e^+ e^-$ colliders, on the other hand,
show good prospects for the search of light
scalars \cite{Wang:2018fcw,Drechsel:2018mgd}. 
The main production channel in the mass and energy range 
that we are interested in is the Higgs-strahlung process
$e^+ e^- \to \phi Z$, where $\phi$ is the
scalar being searched for.
The LEP collaboration has previously performed such
searches~\cite{Barate:2003sz},
which resulted in the $2\,\sig$ excess given by $\mu_{\rm LEP}$.
These searches were limited by the low luminosity of LEP.
However, the ILC, with its much higher luminosity and the possibility 
of using polarized beams, has a substantially higher 
potential to discover the light scalars.
The searches performed at LEP can be divided into two categories:
the 'traditional method', where studies are based on the decay mode
$\phi \to b \bar b$ along with $Z$ decays to $\mu^+ \mu^-$
final states. 
This method introduces certain amount of model dependence 
into the analysis because of the reference to a specific decay mode of $\phi$. 
The more model independent 'recoil technique' used by 
the OPAL collaboration of LEP looked for light states 
by analyzing the 
recoil mass distribution of the di-muon system 
produced in $Z$ decay \cite{Abbiendi:2002qp}. 

In \reffi{fig:2ilc} and \reffi{fig:4ilc} we show previous
bounds from the LEP as well as the projected 
bounds from the ILC searches for light scalars in type~II and type~IV 
N2HDM scenarios respectively. The lines indicating the ILC reach for
a $\sqrt s = 250 \gev$ machine with beam polarizations 
$(P_{e^-}, P_{e^+})$ of $(-80\%,+30\%)$ and an integrated
luminosity of 2000\,$\ifb$ are as evaluated 
in \citere{Drechsel:2018mgd}. The quantity $S_{95}$ used in their
analysis corresponds to an upper limit at the $95\%$ confidence level 
on the cross section times branching ratio generated within the
'background only' hypothesis, where the cross section has been
normalized to the reference SM-Higgs 
cross section and the BRs have been assumed to be as in the SM
(with a Higgs boson of the same mass). Consequently, we take the
obtained limits to be valid for the total cross section times branching ratio.
The colored points shown in \reffi{fig:2ilc} and \reffi{fig:4ilc}
are the points of our scans in the type~II and type~IV scenario
satisfying all the theoretical and experimental constraints.
The plots show that the parameter points 
of our scans can in both cases completely be covered by searches
at the ILC for additional Higgs-like scalars.

Depending on $c_{h_1 V V}$, i.e., the light Higgs-boson production
cross section, the $h_1$ can be produced and analyzed in detail
at the ILC. A detailed analysis of the corresponding experimental
precision of the light Higgs-boson couplings, however,
is beyond the scope of this paper.

%%%%%%%%%%%%%%%%%%%%%%%%%%%%%%%%%%%%%%%%%%%%%%%%%%%%%%%%%%%%%%%%%%%%%%%%%%

\section {Conclusions}
\label{sec:conclusion}

We analyzed a $\sim 3\,\sigma$ excess (local)
in the diphoton decay mode at $\sim 96 \gev$ as reported by
CMS, together with a $\sim 2\,\sigma$ excess (local) in the
$b \bar b$ final state at LEP in the same mass range.
We interpret this possible signal as a Higgs boson in the
2~Higgs Doublet Model with an additional real Higgs singlet (N2HDM),
where this Higgs sector corresponds to the Higgs sectors of the NMSSM or
the  (one-generation case) \mnSSM\ (up to
SUSY relations and
an additional $\cp$-odd Higgs
boson, which is not relevant in our analysis). 

We include all relevant constraints in our analysis. These are
theoretical constraints from perturbativity and the requirement that the
minimum of the Higgs potential is a global minimum. We take into account
the direct searches for additional Higgs bosons from LEP. the Tevatron
and the LHC, as well as the measurements of the properties of the Higgs
boson at $\sim 125 \gev$. We furthermore include bounds from flavor
physics and from electroweak precision data. 

We demonstrate that due to the structure of the couplings of the Higgs
doublets to fermions only two types of the N2HDM, type~II and type~IV
(flipped), can fit simultaneously the two excesses. On the other hand,
the other two types, type~I and type~III (lepton specific), cannot be
brought in agreement with the two excesses. 
Subsequently, we scanned the free parameters in the two favored versions
of the N2HDM, where the results are similar in both scenarios. We find
that the lowest possible values of $\MHp$ above $\sim 650 \gev$ and
$\tb$ just above~1 are favored. The reduced $\chi^2$ from the
Higgs-boson measurements is found roughly in the range 
$0.9 \lesssim  \chi_{\rm red}^2 \lesssim 1.3$. 
Due to the different coupling to leptons in type~II and type~IV, in
general larger values of $\mu_{\rm CMS}$ can be reached in the former,
and the CMS excess can be fitted ``more naturally'' in the type~II N2HDM. 
Incidentally, this is exactly the Higgs sector that is required by
supersymmetric models. 

Finally, we analyzed how the favored scenarios can be tested at future
colliders. The (HL-)LHC will continue the searches/measurements in the
diphoton final state. But apart from that we are not aware of other
channels for the light Higgs boson that could be accessible. 
Concerning the searches for heavy N2HDM Higgs bosons,
particularly interesting are the prospects for charged Higgs bosons.
For the low $\tan\beta$ values favored in our analysis, these
searches have the best potential to discover a new heavy Higgs boson
at the LHC Run III or the HL-LHC.
Also the decay of the heavy neutral Higgs bosons to $t \bar t$
could be promising.

A future $e^+e^-$ collider, such as the ILC, will be able to
produce the light Higgs state at $\sim 96\gev$
in large numbers and consequently study its decay
patterns. Similarly, we demonstrated that
the high anticipated precision in the
coupling measurements of the $125\gev$ Higgs boson
at the ILC (or CLIC, FCC-ee, CepC) will allow to
find deviations in particular in the couplings to massive gauge bosons
if the N2HDM with a $\sim 96 \gev$ Higgs boson is realized in nature.
Here a deviation of more than $2\,\sig$ and $5\,\sig$ at the HL-LHC and
the ILC, respectively, can be anticipated.

We are eagerly awaiting updated analyses from ATLAS and CMS to clarify
the validity of the excess in the diphoton channel.

%%%%%%%%%%%%%%%%%%%%%%%%%%%%%%%%%%%%%%%%%%%%%%%%%%%%%%%%%%%%%%%%%%%%%%%%%%
%\clearpage

\subsection*{Acknowledgements}

We thank 
R.~Santos,
T.~Stefaniak
and
G.~Weiglein
for helpful discussions. M.C.\ thanks D.~Azevedo for discussions regarding
\texttt{ScannerS}.
The work  was supported in part by the MEINCOP (Spain) under 
contract FPA2016-78022-P and in part by the AEI
through the grant IFT Centro de Excelencia Severo Ochoa SEV-2016-0597. 
The work of T.B.\ and S.H. was 
supported in part by the Spanish Agencia Estatal de
Investigaci\'on (AEI), in part by
the EU Fondo Europeo de Desarrollo Regional (FEDER) through the project
FPA2016-78645-P, in part by the ``Spanish Red Consolider MultiDark''
FPA2017-90566-REDC.
The work of T.B. was
funded by Fundaci\'on La Caixa under `La Caixa-Severo Ochoa' international
predoctoral grant.

%%%%%%%%%%%%%%%%%%%%%%%%%%%%%%%%%%%%%%%%%%%%%%%%%%%%%%%%%%%%%%%%%%%%%%%%%%
%\begin{thebibliography}{100}
%
%\end{thebibliography}
\bibliography{n2hdm}

%%%%% CLEAR DOUBLE PAGE!
\newpage{\pagestyle{empty}\cleardoublepage}

%%%%%%%%%%%%%%%%%%%%%%%%%%%%%%%%%%%%%%%%%%%%%%%%%%%%%%%%%%%%%%%%%%%%%%%%%%%

\end{document}

%% file: paperdef.tex
\newcommand\tb{\tan\beta}

\newcommand\ReDiag{\mathop{%
  \raise .5pt\hbox{[}%
  \widetilde{\mathrm{Re}}%
  \raise .5pt\hbox{]}}}
\newcommand\ReOffDiag{\mathop{%
  \raise .5pt\hbox{$\llbracket$}%
  \widetilde{\mathrm{Re}}%
  \raise .5pt\hbox{$\rrbracket$}}}

\newcommand\SW{s_\mathrm{w}}
\newcommand\CW{c_\mathrm{w}}
\newcommand\MW{M_W}
\newcommand\MZ{M_Z}

\newcommand\MHp{M_{H^\pm}}

\newcommand\refeq[1]{Eq.~(\ref{#1})}
\newcommand\refeqs[1]{Eqs.~(\ref{#1})}
\newcommand\refta[1]{Tab.~\ref{#1}}
\newcommand\refse[1]{Sect.~\ref{#1}}

\newcommand\citere[1]{Ref.~\cite{#1}}
\newcommand\citeres[1]{Refs.~\cite{#1}}

\newcommand{\SM}{\mathrm{SM}}

\newcommand{\mnSSM}{\ensuremath{\mu\nu\mathrm{SSM}}}

\newcommand{\CP}{{\cal CP}}
\newcommand{\cp}{{\CP}}

\newcommand{\tev}{\,\, \mathrm{TeV}}
\newcommand{\gev}{\,\, \mathrm{GeV}}

\newcommand{\Hpm}{H^\pm}

\newcommand\fb{\ensuremath{\mbox{fb}}}

\newcommand\ifb{\ensuremath{\fb^{-1}}}

\newcommand{\br}{\text{BR}}

\newcommand{\sig}{\sigma}

\def\reffi#1{\mbox{Fig.~\ref{#1}}}
\def\reffis#1{\mbox{Figs.~\ref{#1}}}

\def\ga{\gamma}

\definecolor{Orange}{rgb}{0.8,0.5,0.}
\definecolor{Purple}{rgb}{0.5,0.,0.5}
\definecolor{Lightblue}{cmyk}{0.9,0.1,0.1,0.3}
\definecolor{dgelborange}{cmyk}{0.,0.3,0.5, 0.}
\definecolor{Lila}{rgb}{0.5,0.,1}